%% file: main.tex
\documentclass{article}
\usepackage[final,sglblindworkshop]{neurips_2026}
\workshoptitle{World Models for High-Stakes Health}
\usepackage[utf8]{inputenc}
\usepackage[T1]{fontenc}
\PassOptionsToPackage{hyphens}{url}\usepackage{hyperref}
\usepackage{url}
\usepackage{booktabs}
\usepackage{amsfonts}
\usepackage{amsmath}
\usepackage{nicefrac}
\usepackage{microtype}
\usepackage{xcolor}
\usepackage{graphicx}
\usepackage{multirow}
\usepackage{siunitx}
\usepackage{caption}
\usepackage{subcaption}
\usepackage{enumitem}
\hypersetup{colorlinks=true, linkcolor=blue!60!black, citecolor=blue!60!black, urlcolor=blue!60!black}
\newcommand{\IS}{\mathrm{IS}}
\newcommand{\blind}{\textsc{blind}}
\newcommand{\full}{\textsc{full}}
\newcommand{\gbt}{\textsc{gbt}}
\newcommand{\lm}{\textsc{lm}}
\newcommand{\ci}[2]{{\footnotesize[$#1$,\,$#2$]}}
\graphicspath{{figures/}{./}}

\title{Does Your Trial Simulator Know\\ What the Drug Does?\\[3pt]
{\large An Intervention-Blind Test of Interventional Validity\\ on 7{,}207 Registered Trials}}

\author{%
  Alicem Koyun\\
  Westcliff University\\
  \texttt{\authoremail}
}

\newcommand{\authoremail}{alicemkoyun@gmail.com}
\newcommand{\repourl}{https://github.com/alicemkyn/ivbench}
\newcommand{\zenododoi}{10.5281/zenodo.23521618}
\makeatletter\newcommand{\doiclause}[1]{\if\relax\detokenize\expandafter{\zenododoi}\relax\else#1\fi}\makeatother
\begin{document}

\maketitle

\begin{abstract}
Patient world models are increasingly proposed as simulators of clinical trials, and recent surveys organise their capabilities as a ladder that rises from temporal prediction, through action-conditioned prediction, to counterfactual rollouts that can support decisions. Position on that ladder is currently assigned by reading papers rather than by measurement. We turn the counterfactual rung into a measured quantity. From 30{,}119 completed phase~2/3 trials with results posted to ClinicalTrials.gov we construct 9{,}996 audited within-trial contrasts (treatment-arm outcome minus comparator-arm outcome on the primary endpoint) from 7{,}207 trials, and we score a simulator not on how well it predicts a held-out arm but on how much of its accuracy \emph{depends on knowing which intervention was tested}. The instrument is an intervention-blind reference: the same model, trained on the same trials, with every intervention-identifying field withheld. \emph{Interventional skill} is the error the full model removes relative to that reference. On a temporal split (results posted before/after 2023; 878 test contrasts from 655 trials) a registry-trained gradient-boosting simulator removes a fifth of the error of a trivial baseline, but its interventional skill is $-0.12$ (95\% CI $-0.21$ to $-0.05$, five seeds): withholding the intervention makes it better, and no alternative featurisation, stratum, temporal cut or same-molecule retrieval yields an interval above zero, while the sampling-noise ceiling of the task is $0.91$ on skill against the global mean ($0.87$ relative to the blind reference). A frontier language model reading only the registered protocol has interventional skill $0.20$ \ci{0.08}{0.31}; a smaller model of the same family and an open-weight 70B model have $0.08$ \ci{-0.01}{0.17} and $0.08$ \ci{0.00}{0.15}, at the limit of detection. Perturbation tests show the large model uses the intervention as such: swapping the arms reverses the sign of its prediction in 86\% of contrasts and substituting an unrelated drug is worse than withholding the drug. The blind reference is itself not drug-blind: from the blind prompt a language model recovers the withheld drug class 45\% of the time (chance 2\%), so interventional skill is a conservative lower bound on what the intervention contributes. Where exposure to posted results can be excluded (217 contrasts completed after an open-weight model's training cutoff) its protocol-only interventional skill is $0.01$ \ci{-0.10}{0.12}, too imprecise to confirm or exclude memorisation. A reliability layer shows that ranking by predicted error lowers it but that calibrated error control does not transfer under temporal shift. Benchmark, code, prompts and per-call model outputs are released.
\end{abstract}

\section{Introduction}
\label{sec:intro}

A clinical trial is the one place in medicine where a counterfactual is actually observed: two exchangeable groups, one intervention, one contrast. That makes trial simulation the most falsifiable application of a \emph{patient world model}, a generative or reasoning system that learns from trajectories, protocols and interventions and rolls forward what happens under a treatment it has not seen applied to this population; and it makes current evaluation practice inadequate. Simulators are scored on held-out prediction, but a model can predict a held-out arm well by learning what a responder rate in ulcerative colitis usually looks like or how large a placebo response an enriched population produces, neither of which involves the treatment. What matters for decision support is the \emph{difference} between arms, and for that quantity held-out accuracy is not evidence of understanding.

The recent survey of world models in healthcare~\citep{qazi2025worldmodels} makes this concern explicit. It organises the field on a four-rung ladder of temporal prediction (L1), action-conditioned prediction (L2), counterfactual rollouts (L3) and planning (L4), finds that electronic-health-record systems cluster at L1, and names \emph{interventional validity} as the property the field cannot yet check: simulated futures ``may be realistic but wrong about treatment effects''. Rungs there are assigned by reading the papers; our own screen of 447 clinically grounded NeurIPS 2025 workshop papers (Appendix~\ref{app:screen}) finds 26 that model a trajectory or trial arm and none that both represents the intervention and reports, at abstract level, validation against a real trial arm.

\paragraph{This paper.} We make interventional validity a measured quantity. The construction has three parts.

\begin{enumerate}[leftmargin=*, itemsep=1pt]
\item \textbf{A benchmark of real counterfactuals.} From every completed phase~2/3 interventional trial with results posted to ClinicalTrials.gov we extract arm-level primary-endpoint results and form within-trial contrasts, treatment arm minus comparator arm, as a risk difference or a standardised mean difference. The audited release holds 9{,}996 contrasts from 7{,}207 trials, with a temporal split on the results-posting year (Section~\ref{sec:benchmark}).
\item \textbf{An intervention-blind control.} For any simulator we train a second copy on identical trials with every intervention-identifying field withheld but endpoint, condition, eligibility, design, population and comparator type intact. \emph{Interventional skill} is one minus the ratio of the full model's mean squared error to the blind model's (Section~\ref{sec:is}), the trial-level analogue of the hypothesis-only and question-only baselines of NLI and VQA~\citep{poliak2018hypothesisonly, gururangan2018artifacts, goyal2017vqa}: if a prediction survives removal of the treatment, it was not a prediction of a treatment effect.
\item \textbf{Systems and tests of the measurement.} We apply the control to a registry-trained gradient-boosting simulator and to three protocol-reading language models, and test the control itself: how much of the drug the blind reference still implies, whether the models respond to the intervention as an intervention (arm swap, drug substitution), and whether exposure to posted results explains the result (Section~\ref{sec:results}).
\end{enumerate}

\paragraph{Contributions.} (i)~A public, audited benchmark of 9{,}996 real within-trial contrasts with construction code, data card, splits and every result table regenerable by one script.\footnote{Code and data: \url{\repourl}\doiclause{; archived release: \href{https://doi.org/\zenododoi}{doi:\zenododoi}}.} (ii)~Interventional skill, an intervention-blind measurement that operationalises the L3 rung of~\citet{qazi2025worldmodels}, to be reported alongside held-out accuracy as a majority-class baseline is for a classifier, together with the finding that its blind reference is not drug-blind and the measurement is therefore a lower bound. (iii)~The finding that registry-trained simulators have no interventional skill while a frontier language model reading the protocol has a modest, perturbation-consistent amount and smaller models sit at the limit of detection. (iv)~A reliability layer showing that confidence ranking helps but calibrated error control does not survive temporal shift.

\section{Related work}
\label{sec:related}

\paragraph{Predicting trial success is not predicting effect size.} Machine learning for trials has predicted whether a trial \emph{succeeds}: approval prediction~\citep{lo2019approval, siah2021novartis}, HINT~\citep{fu2022hint}, SPOT~\citep{wang2023spot}, Trial2Vec~\citep{wang2022trial2vec}, LLM systems that read protocols~\citep{zheng2025lifted, zhang2025mexactp, gao2024lint, reinisch2024ctpllm, zhang2025cladmop, zheng2026deepimagine} and benchmarks built on the success label~\citep{chen2025trialbench, gao2026cto}; a 2025 scoping review finds effect-size prediction essentially absent~\citep{teodoro2025scoping}. Success confounds effect magnitude with power, endpoint choice and sponsor strategy, and none of these systems is scored against a treatment-blind reference. LLMs match HINT on success yet miss negative trials~\citep{jin2024canai}; design features add signal beyond drug and indication~\citep{howard2026pos}. The only prior work predicting posted arm-level results from registrations is \citet{hu2026sae}, who predict which arm has more serious adverse events: safety, direction, no blind control.

\paragraph{Digital twins predict the control arm.} Industry ``digital twins'' are prognostic models of the control-arm outcome~\citep{fisher2019alzheimer, walsh2020digitaltwins}, deployed through PROCOVA, a covariate adjustment that is unbiased \emph{because} the prognostic score is estimated without the treatment~\citep{schuler2022procova, ema2022procova, vidovszky2024digitaltwins}. External and synthetic control arms make the stronger claim, and every guidance conditions it on comparability checks against concurrent controls~\citep{thorlund2020synthetic, ghadessi2020roadmap, viele2014historical, elvatun2025synthetic, fda2023externallycontrolled}. LLM-based trial simulation~\citep{wang2024twingpt, makarov2026twinweaver, xu2024synthetictrial, yue2024clinicalagent, jin2024trialgpt} is evaluated on record fidelity, success labels or matching; none reports whether the simulated trial reproduces the posted contrast or whether the output changes when the intervention is hidden.

\paragraph{Causal evaluation lacks a null-by-construction reference.} CATE benchmarking~\citep{hill2011ihdp, dorie2019acic, shimoni2018benchmarking, neal2020realcause, parikh2022credence} has a matching critique: rankings are artefacts of simulation choices and causal models need interventional evaluation on empirical data~\citep{curth2021cate, mahajan2022modelselection, gentzel2019interventional, curth2024cpt, feuerriegel2024causalml}. Missing is a negative control~\citep{lipsitch2010negative} applied to a learned model: the hypothesis-only baselines of NLI~\citep{poliak2018hypothesisonly, gururangan2018artifacts}, question-only priors in VQA~\citep{goyal2017vqa} and shortcut learning generally~\citep{geirhos2020shortcut}. Target trial emulation~\citep{hernan2016target, hernan2022target, hansford2023reporting} and RCT-DUPLICATE~\citep{franklin2019process, franklin2020duplicate, wang2023emulation} judge one bespoke observational study against one RCT; here a learned model is judged on thousands, against a reference that hides the treatment. Meta-research explains why design features carry skill: concealment and blinding~\citep{wood2008bias}, trial size~\citep{dechartres2013samplesize, pereira2012large}, control-group event rate~\citep{schmid1998controlrate} and a predictable placebo response~\citep{furukawa2016placebo, tuttle2015placebo, enck2013placebo} predict effect estimates, while the prior that a new treatment beats an established one is near even~\citep{djulbegovic2012new} and expert forecasts are modest~\citep{atanasov2020wisdom, benjamin2017preclinical}; the intervention-specific component is what a simulator exists to add.

\paragraph{World models, reliability, registry.} Healthcare world models~\citep{ha2018worldmodels, qazi2025worldmodels, chen2026medicalworldmodels, kraljevic2024foresight, renc2024ethos, waxler2025comet, shmatko2025delphi, yang2025mewm, mu2026ehrworld, xu2025meddreamer} are evaluated predictively~\citep{wornow2023ehrshot, wornow2023shaky}. Conformal methods~\citep{vovk2022algorithmic, lei2018distributionfree, angelopoulos2021gentle, romano2019cqr, tibshirani2019covariate, vazquez2022conformal}, selective prediction~\citep{geifman2017selective, chen2024uncertainty} and abstention~\citep{kompa2021uncertainty, vancalster2019calibration} supply the reliability layer. The CT.gov results database~\citep{zarin2011results, tse2018problems, tasneem2012aact} and its incomplete, sponsor-dependent compliance~\citep{zarin2016finalrule, prayle2012compliance, anderson2015compliance, devito2020compliance, hartung2014discrepancies, riveros2013timing} bound what a benchmark built on it can claim.

\section{A benchmark of real within-trial contrasts}
\label{sec:benchmark}

\begin{figure}[t]
\centering
\includegraphics[width=\linewidth]{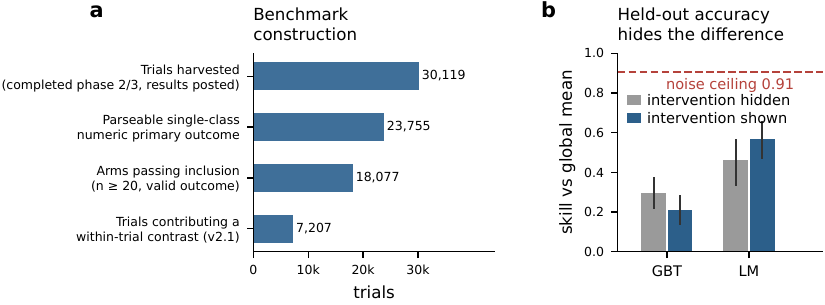}
\caption{\textbf{Held-out accuracy does not separate the systems; the blind control does.} (a)~Benchmark construction. (b)~Skill against the global training mean on the 878 binary test contrasts with the intervention shown or withheld; GBT = registry-trained gradient boosting (five-seed mean), LM = large language model (protocol-only prompt, single run); whiskers 95\% trial-bootstrap intervals; dashed line noise ceiling.}
\label{fig:benchmark}
\end{figure}

\paragraph{Source and inclusion.} We queried the ClinicalTrials.gov v2 API on 9 September 2026 for completed phase~2, 2/3 or 3 interventional studies with posted results (30{,}119). A study enters if its first primary outcome has a single reported class, a numeric parameter type (mean, least-squares mean, median or geometric mean; count or number of participants) and per-group analysed denominators (23{,}755). Arms with fewer than 20 analysed participants or a non-finite outcome are dropped, leaving 40{,}450 arm records from 18{,}077 trials (Figure~\ref{fig:benchmark}a).

\paragraph{Arm records and roles.} Each arm record carries the endpoint (title, description, timeframe, unit), the outcome on a common scale, the analysed $n$, baseline mean age and fraction female where reported, the arm's protocol label and description, its intervention names, types and descriptions, and its results-group title and description (join details in Appendix~\ref{app:construction}; 98.7\% of contrasts carry substantive intervention text). Each arm is classified as \emph{inert control} (placebo, sham, no intervention), \emph{active comparator}, or \emph{treatment}, from the registered arm type where it can be recovered and the group text otherwise; the comparator's nature is itself predictive, so the two control kinds are kept distinct.

\paragraph{Contrasts and the pairing audit.} Within a trial and outcome, each treatment arm is paired with the largest inert-control arm if one exists, otherwise the largest active comparator (up to three contrasts per trial). Binary endpoints yield the risk difference $\Delta = p_T - p_C$; continuous endpoints the standardised mean difference $d = (\bar y_T - \bar y_C)/s_{\text{pooled}}$, with $s$ reconstructed from the reported SD, SE or 95\% interval; contrasts with $|d| > 5$ are removed. The submitted version (v2; 11{,}274 contrasts) paired arms by text alone; a model-assisted audit of 100 random contrasts found every value correctly re-derived but only 76 pairings valid, and the 24 flagged cases, reviewed by hand, showed three failure modes: a same-molecule comparator at another dose, arms from different stages of a staged or crossover design, and a placebo-labelled double-dummy experimental arm taken as the control. The released version (v2.1) adds three pairing rules (same-molecule exclusion, concurrency, investigational-arm preference; Appendix~\ref{app:construction}) over re-fetched registered arm types; they drop 1{,}510 v2 contrasts, re-pair 375 and add 240. A fresh audit of 100 contrasts finds 96 valid pairings \ci{0.90}{0.98}; all 125 new or re-paired test contrasts were audited individually and 8 removed by hand. The final table holds 9{,}996 contrasts from 7{,}207 trials (4{,}014 risk differences, 5{,}982 standardised differences; 7{,}669 inert-control, 2{,}327 active-comparator); v2 is kept as a sensitivity set. The repair matters: the removed contrasts were those on which the language model gained most from the intervention (its posted-record $\IS$ falls from $0.39$ on v2 to $0.27$ on v2.1), because in a same-molecule dose comparison the ``intervention'' names the dose and a posted group description often names the outcome.

\paragraph{Splits and strata.} The primary split is temporal on the results-posting year: 7{,}789 contrasts posted before 2023 for training, 2{,}207 posted in 2023 or later for testing (binary: 3{,}136 train, 878 test from 655 trials; continuous: 4{,}653 and 1{,}329 from 978). Six therapeutic-area strata support leave-one-area-out evaluation; design flags (randomised parallel-group allocation, 96\% of binary test contrasts; participant-level outcome unit) define sensitivity subsets (Tables~\ref{tab:design} and~\ref{tab:designcont}).

\paragraph{How much is predictable at all?} A reported contrast is the true contrast plus sampling noise. From the arm proportions and sizes, the mean sampling variance of a binary test contrast is $0.0047$ against a total variance of $0.0499$: reliability $0.906$ \ci{0.885}{0.922}, the ceiling on skill against a global mean. Relative to the blind registry model's residual error the attainable interventional skill is at most $0.87$, relative to the large language model's blind prompt $0.82$; every model below sits far beneath these, so a null cannot be attributed to noise.

\section{Interventional skill}
\label{sec:is}

\paragraph{Skill.} For a predictor $\hat\Delta$ and a reference $\tilde\Delta$ on a test set, $\mathrm{skill}(\hat\Delta \mid \tilde\Delta) = 1 - \mathrm{MSE}(\hat\Delta)/\mathrm{MSE}(\tilde\Delta)$: the fraction of the reference's squared error the predictor removes. Against the global training mean this is the usual $R^2$-style number. Confidence intervals are percentile bootstraps over \emph{trials}, because contrasts from one trial share arms; five-seed means use one shared set of resamples.

\paragraph{The blind reference.} The reference we care about is a copy of the same model, trained on the same trials with the same procedure and the same endpoint text, condition list, eligibility criteria, design fields, population covariates, therapeutic area and comparator type, and denied every field that identifies the intervention (intervention names, types and descriptions, protocol arm labels and descriptions, results-group titles and descriptions). Call these \blind\ and \full; \blind$^{-}$ is the same reference without comparator type. \emph{Interventional skill} is
\[
\IS \;=\; \mathrm{skill}(\hat\Delta_{\full} \mid \hat\Delta_{\blind}) \;=\; 1 - \frac{\mathrm{MSE}(\hat\Delta_{\full})}{\mathrm{MSE}(\hat\Delta_{\blind})} .
\]
A simulator that models the effect of the treatment should have $\IS > 0$ whenever the withheld fields are its only route to the intervention; Section~\ref{sec:blind} shows they are not entirely, so $\IS$ is a lower bound. A model that has learned only endpoint scale, population prognosis and comparator type has $\IS = 0$ by construction, however high its skill against the global mean; $\IS<0$ arises when the intervention fields add variance without information.

\paragraph{Why this is the right control, and what it misses.} The global mean credits the model for endpoint scale; shuffling the intervention across trials destroys the legitimate intervention--endpoint co-occurrence; withholding the field with everything else intact isolates the increment that knowing the treatment buys, as the hypothesis-only baseline does in NLI~\citep{poliak2018hypothesisonly}. It does not isolate it completely: condition, endpoint and eligibility text imply the drug class (a biologic-na\"ive ulcerative-colitis trial with a remission endpoint is a trial of an advanced therapy), so the blind reference already carries class-level knowledge and $\IS$ measures the increment of the specific drug \emph{beyond what the rest of the protocol implies}. Section~\ref{sec:blind} quantifies this and shows that $\IS$ is a conservative lower bound.

\paragraph{Reading $\IS$ with a language model.} For a prompted system the control is the same prompt with the intervention fields replaced by a withheld marker and intervention-derived tokens removed from the retained text (Appendix~\ref{app:llm}). The prompt is built from registered protocol fields only (``protocol-only'': the condition is defined before the trial reads out); the submitted version's prompt also carried the posted results-group labels and descriptions (``posted-record''), which can leak outcome vocabulary, and is kept as a second row. Two further confounds, pretraining exposure to the trial's results and redaction leakage, are addressed by stratifying trials on completion, posting and publication dates relative to a model's documented training cutoff and by auditing every blind prompt.

\section{Experiments}
\label{sec:results}

\subsection{Systems}

\paragraph{Registry-trained simulator.} The main system is a histogram gradient-boosting regressor (Appendix~\ref{app:features}) over TF-IDF text reduced to 120 SVD components, ordinal-encoded categoricals and raw numerics; \full\ text is endpoint plus intervention fields, \blind\ text endpoint fields only. Both are trained on the 3{,}136 binary (4{,}653 continuous) training contrasts over five seeds. We also report an \emph{intervention-only} model and six alternative featurisations (Section~\ref{sec:gbt}).

\paragraph{Protocol-reading language models.} Three language models are prompted once per contrast with the protocol fields rendered as text and asked for a point estimate and a 90\% interval for the risk difference as strict JSON: a frontier model (Claude Sonnet~5, ``large''), a smaller model of the same family (Claude Haiku~4.5, ``small''), and an open-weight model with a documented December 2023 knowledge cutoff (Llama~3.1 70B Instruct, unquantised BF16 on a dedicated deployment, temperature 0). Condition~A (\full) includes the arms' labels, intervention names and descriptions and protocol arm descriptions; condition~B (\blind) replaces each of those lines with a withheld marker and passes the retained fields through a redaction that removes intervention-derived name tokens. Both conditions were run on all 878 binary test contrasts with all three models under protocol-only and posted-record prompts, and on the 1{,}329 continuous contrasts with the large model (posted-record prompt, SMD-scale header); every prompt parsed (Appendix~\ref{app:llm}). The redaction audit finds no surviving intervention-derived token in any protocol-only blind prompt and four ordinary English words across the posted-record ones; class-level cues in eligibility text remain available to the blind condition, which makes the reference stronger and the estimate conservative (Section~\ref{sec:blind}).

\subsection{Held-out accuracy looks fine for everyone}

Against the global training mean (Table~\ref{tab:main}, Figure~\ref{fig:benchmark}b), the registry-trained \full\ model removes 21\% of baseline error on risk differences and 29\% on standardised differences; the large language model removes 57\% \ci{0.47}{0.66}, the two smaller models 35\% and 40\%. Read as a leaderboard, every system predicts treatment effects.

\input{tables/main}

\subsection{The registry-trained simulator has no interventional skill}
\label{sec:gbt}

On v2.1 the registry model's interventional skill is $-0.12$ \ci{-0.20}{-0.05} over five seeds (Table~\ref{tab:main}): adding every intervention field lowers held-out skill from $0.30$ to $0.21$. The change from the submitted version ($-0.05$ \ci{-0.13}{0.02} on v2) is in the reference, not in the full model: \full\ skill is $0.21$ on both versions, while \blind\ rises from $0.25$ to $0.30$ because the repaired comparator labels make comparator type more informative (its gain over \blind$^{-}$ grows from $0.02$ to $0.04$; Table~\ref{tab:seeds}). The negative sign is mostly a cost of placing intervention text in the endpoint's shared SVD space: with separate text blocks, sparse ridge or PubMedBERT embeddings the estimate lies between $-0.05$ and $-0.01$, and no architecture's interval extends above $0.03$ (Table~\ref{tab:arch}). A two-stage model, which can add to the \blind\ prediction only a cross-validated and shrunken correction from intervention text and so stays close to zero by construction, gives $0.00$ \ci{0.00}{0.01}. We read the row as no detectable interventional skill, and treat $\IS<0$ as a possible outcome of the instrument rather than as a separate finding.

\begin{figure}[t]
\centering
\includegraphics[width=\linewidth]{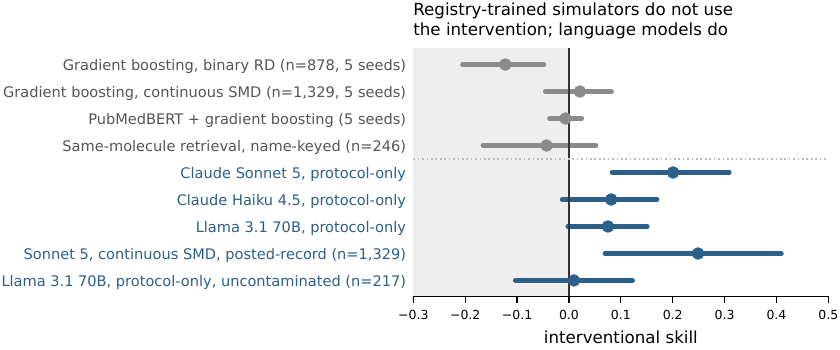}
\caption{\textbf{Interventional skill by system and stratum (v2.1).} Whiskers are 95\% trial-bootstrap intervals; $n$ = test contrasts. Grey labels: registry-trained models; blue labels: language models with protocol-only prompts (continuous row posted-record), and Llama~3.1 70B on the uncontaminated stratum (trials completed, posted and published after its training cutoff).}
\label{fig:forest}
\end{figure}

\paragraph{The null survives every alternative explanation we could test} (Figure~\ref{fig:forest}; Appendix~\ref{app:tables}). \emph{Endpoint family:} 1{,}329 continuous contrasts, $\IS = 0.02$ \ci{-0.05}{0.08}. \emph{Temporal cut:} retrained splits at 2021, 2022 and 2024 give $-0.04$, $-0.06$ and $-0.11$. \emph{Comparator, phase and text strata:} retrained within placebo-controlled ($n=644$), active-controlled (234), phase-3 (445) and arm-matched rich-text (697) contrasts, $\IS\in[-0.10, -0.01]$; phase-2-only trials give $0.08$ \ci{-0.05}{0.17}. \emph{Arm-level task:} predicting each of 4{,}177 test arms' own event proportion, \blind\ ($0.245$) again beats \full\ ($0.212$). \emph{Seen molecules do not help:} on the 286 test contrasts whose intervention name appears in training, $\IS$ is $-0.23$ \ci{-0.39}{-0.08}, more negative than on the 592 unseen ($-0.10$ \ci{-0.19}{-0.01}; Table~\ref{tab:seen}, Figure~\ref{fig:probes}b). \emph{Feature groups} (Figure~\ref{fig:probes}a): over a design-only base, endpoint text adds $0.13$ \ci{0.04}{0.22} and intervention text $-0.01$ \ci{-0.08}{0.06}; the intervention-only model reaches $0.05$ against the training mean, so the fields carry information, none of it beyond what endpoint, population and comparator already give.

\paragraph{Positive control: is interventional skill attainable from the registry at all?} If the treatment's effect is anywhere in public data, it is in earlier trials of the same molecule. For the 246 test contrasts (201 trials) with a training contrast of the same normalised intervention name and comparator type, predicting the blind value plus the shrunk mean residual of the matched trials gives $\IS = -0.04$ \ci{-0.17}{0.05}; the \full\ model on the same contrasts gives $-0.22$ (Table~\ref{tab:mol}). The registry does not contain, in a form these models can extract, the increment that knowing the drug should provide.

\subsection{Language models have interventional skill, in proportion to scale}
\label{sec:lm}

Applied to the language models, the control gives a different answer. Withholding the intervention drops the large model's skill from $0.570$ to $0.462$: $\IS = 0.20$ \ci{0.08}{0.31} with protocol-only prompts, $0.27$ \ci{0.16}{0.37} with the submitted version's posted-record prompts (paired difference $0.07$ \ci{-0.03}{0.17}). Its blind skill exceeds the registry model's full skill: it reads eligibility criteria and endpoint definitions better than TF-IDF does, so held-out accuracy ranks it first for a reason unrelated to the drug. The smaller models sit at the limit of detection: Haiku~4.5 $0.08$ \ci{-0.01}{0.17}, Llama~3.1 70B $0.08$ \ci{0.00}{0.15}. The large model's skill reappears on continuous endpoints ($0.25$ \ci{0.07}{0.41}, posted-record), on the randomised parallel-group subset ($0.22$) and on contrasts with a participant-level outcome unit ($0.19$; Table~\ref{tab:design}); a second run of its blind condition moves $\IS$ by $0.02$ \ci{-0.02}{0.05} (Table~\ref{tab:lmrevision}).

\paragraph{Does the model use the intervention as an intervention?} Two perturbations of the large model's posted-record prompt test this (Table~\ref{tab:lmrevision}). \emph{Arm swap:} exchanging the arms reverses the sign of the prediction in 86\% of the 742 contrasts with a non-trivial original prediction (correlation $0.95$ with the negated original); the registry model, whose bag-of-words features are nearly order-invariant, reverses 7\%. \emph{Pre-registered negative control:} replacing the treatment by an unrelated marketed drug (loratadine in cardiometabolic trials, ezetimibe elsewhere) shrinks the predicted effect ($0.035$ \ci{0.020}{0.049}; H1 supported), more so in placebo-controlled trials ($0.073$; H2's mean criterion supported, its slope criterion of $<0.5$ not met at $0.59$ \ci{0.50}{0.69}), and in active-comparator trials with a predicted benefit turns it towards harm ($-0.12$ \ci{-0.18}{-0.07}; H3 supported). The wrong drug is worse than no drug: $\mathrm{skill}(\mathrm{NC}\mid\blind) = -0.87$ \ci{-1.34}{-0.47}.

\paragraph{Exposure to the results.} The Claude models' documented training cutoffs leave no usable post-cutoff stratum (no binary test contrast for Sonnet~5; 5 for Haiku~4.5), so we test exposure with the open-weight model. A contrast is uncontaminated for Llama~3.1 70B when its trial's primary completion and first results posting both fall on or after 1 January 2024 and no linked results publication predates that date (217 contrasts, 152 trials; Table~\ref{tab:contamination}). In the uncontaminated stratum the posted-record interventional skill is unchanged from the full test set ($0.09$ \ci{-0.01}{0.20} against $0.09$ \ci{0.02}{0.16}), so any gain here cannot come from a memorised result, while the protocol-only value is $0.01$ \ci{-0.10}{0.12}, against $0.10$ \ci{0.01}{0.19} in the contaminated complement (difference $-0.09$ \ci{-0.24}{0.06}). The two instruments say different things, and at this sample size the comparison can neither confirm nor exclude a drop of the size the point estimates suggest (finer splits in Appendix~\ref{app:tables} are equally imprecise). The posted-record result is consistent with interventional skill being attainable without exposure to the results, though its interval touches zero; we make no memorisation claim for the Claude models.

\subsection{How blind is \blind?}
\label{sec:blind}

The blind reference withholds the intervention's name, not its implications. Given only the blind prompt, Haiku~4.5 guesses the withheld treatment's drug class correctly for 45\% \ci{0.41}{0.49} of test contrasts, its mechanism for 45\% and the exact agent for 21\%, against 2\%, 2\% and 1\% when the same guesses are scored against a randomly permuted contrast's treatment (Table~\ref{tab:blindness}; a Haiku~4.5 judge scored the guesses). The registry model's blind text behaves the same way: for test contrasts whose molecule occurs in training, the nearest training contrast by blind TF-IDF text is a same-molecule contrast 23\% of the time (chance $0.3\%$), 19\% in the 120-dimensional representation the model uses. Both references already encode the drug class and often the drug. $\IS$ is the increment of knowing the \emph{specific} agent beyond what the protocol implies about its class, so it understates the intervention's total contribution and is a conservative lower bound: the registry model's zero means ``nothing beyond what \blind\ already carries'', and the language model's $0.20$ is an increment on a reference that has already placed the trial in a drug class. Any intervention-blind baseline should come with this audit of how much of the withheld variable the remaining inputs reconstruct.

\subsection{Reliability: coverage, ranking and error control}
\label{sec:reliability}

\begin{figure}[t]
\centering
\includegraphics[width=\linewidth]{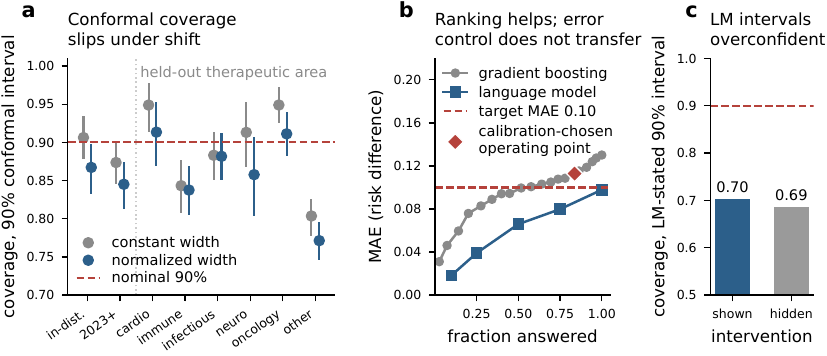}
\caption{\textbf{Reliability layer (v2.1).} (a)~Coverage of 90\% cross-fitted conformal intervals for the registry model, in-distribution, under temporal shift and with each therapeutic area held out. (b)~Selective prediction on the shifted test set, ranking by cross-fitted error scale (registry model) or self-stated interval width (language model, protocol-only); the marked point is the registry operating point whose threshold was chosen on calibration data for a target MAE of $0.10$. (c)~Coverage of the large language model's own 90\% intervals.}
\label{fig:reliability}
\end{figure}

Neither system's point prediction is reliable enough to act on; they fail differently (Figure~\ref{fig:reliability}). Cross-fitted conformal 90\% intervals for the registry model~\citep{lei2018distributionfree, angelopoulos2021gentle}, constant-width and heteroscedastic, cover $0.906$ \ci{0.879}{0.933} and $0.867$ in-distribution but $0.874$ \ci{0.847}{0.899} and $0.845$ under the temporal shift, and $0.77$ to $0.95$ with each therapeutic area held out (Table~\ref{tab:coverage}). They are wide (mean width $0.60$ in-distribution). The language models' self-stated 90\% intervals are less than half as wide ($0.23$) and cover $0.70$ (intervention shown) and $0.69$ (hidden) for the large model and $0.58$ to $0.63$ for the smaller ones: confidently wrong~\citep{vancalster2019calibration}.

Ranking by predicted error lowers it; calibrated error control does not transfer. Answering only the quarter of shifted-set contrasts the registry model's cross-fitted error scale marks most confident cuts MAE from $0.130$ to $0.083$. But a threshold chosen on calibration data to hold MAE at $0.10$ realises $0.106$ in-distribution and $0.113$ under shift, and at a target of $0.08$ realises $0.095$ and $0.100$ (Table~\ref{tab:op}; the submitted version chose thresholds on the evaluated set, which meets any target by construction). Ranking the large language model by its stated interval width, the most confident quarter has MAE $0.039$ against $0.098$ overall (Table~\ref{tab:lmsel}). Coverage without width, width without coverage, or an answered fraction at a target chosen after the fact all hide the operating point that matters: the fraction answerable at an error stated \emph{before} the shift.

\section{Discussion}
\label{sec:discussion}

\paragraph{What the measurement changes.} Every accuracy number in Table~\ref{tab:main} would, under current practice, be reported as evidence that a model predicts treatment effects. The blind control shows that for the registry-trained simulator all of it, and for the large language model about two-thirds, is prediction of endpoint, population and comparator: what meta-research says the registry contains~\citep{wood2008bias, schmid1998controlrate, furukawa2016placebo}.

\paragraph{On the L3 rung.} \citet{qazi2025worldmodels} place decision-supporting counterfactual rollouts at rung L3 and ask for interventional validity to be measured; interventional skill is one such measurement, costing a second training run or prompt. Patient-timeline world models need patient-level EHR that no public benchmark pairs with a posted trial result, but any system that emits an arm-level prediction from a protocol can be scored. By this measure a registry-trained simulator is at L1--L2; a frontier language model has a foothold on L3 ($0.20$ from the protocol alone, far from the attainable $0.82$, overconfident); smaller models are at the limit of detection. We suggest that $\IS$ be reported alongside held-out accuracy, as classifiers report a majority-class baseline, with a blindness audit of the reference. The plausible source of the language model's skill is pretraining exposure to pharmacology and published effects of drug \emph{classes}, which the registry cannot supply: it records what was tested, not what the drug does. That suggests a route for registry-trained simulators (class-level and mechanism priors as inputs) and a caution: real interventional skill with $0.70$ coverage on 90\% intervals is more dangerous in a decision loop than none.

\paragraph{Limitations.}\label{sec:limitations} The benchmark inherits the registry's failure modes: results are posted for a minority of due trials, with sponsor type the strongest predictor~\citep{devito2020compliance, anderson2015compliance}, so the contrasts are a selected sample; endpoints are harmonised into two effect families; covariates are trial-level means. Arm pairing is rule-based and audited by sampling (96 of 100 valid in v2.1); 7{,}690 result groups lack a recoverable registered arm type, and the audits were model-assisted with flagged cases reviewed by the author. Interventional skill is a \emph{necessary} condition for interventional validity, not a sufficient one: positive $\IS$ could exploit a correlate of the intervention rather than its mechanism, and says nothing about individual-level counterfactuals. The exposure test is underpowered (152 uncontaminated trials; interval $\pm0.11$), concerns an open-weight model at temperature 0 while the Claude runs used default sampling, and is not evidence about the Claude models, for which no post-cutoff stratum exists. Two tables of the submitted version could not be reproduced from preserved code and are reported afresh or flagged provisional (Appendix~\ref{app:tables}).

\section{Responsible-use statement}
\label{sec:responsible}

This work evaluates whether models can predict randomised-trial results from protocols. Its negative result argues \emph{against} using registry-trained simulators to replace or shrink control arms; its positive result for language models is modest ($0.20$), with intervals covering $0.58$--$0.70$ at nominal $0.90$ and no error control under shift. The main misuse risk is the inverse one, quoting held-out accuracy to justify a smaller trial or an external control arm; any downstream use should report interventional skill against an audited intervention-blind reference, coverage \emph{and} width of intervals, and the answered fraction at a pre-fixed error target; no such model should inform a design decision without a prespecified comparison to concurrent controls~\citep{fda2023externallycontrolled}.

\bibliographystyle{plainnat}
\bibliography{references}

\newpage
\input{_appendix1}

\input{_appendix2}

\end{document}

%% file: tables/main.tex
\begin{table}[t]
\centering
\small
\caption{\textbf{Skill against the global training mean and interventional skill, v2.1 temporal test split.} Binary: 878 contrasts / 655 trials; continuous: 1{,}329 / 978. \gbt: five-seed mean, 95\% interval from one shared set of 1{,}000 trial resamples; language models (large = Claude Sonnet~5, small = Claude Haiku~4.5; Llama~3.1 70B, open weights): single runs, 2{,}000 trial resamples. Protocol-only prompts hold registered fields only; posted-record prompts (the submitted version's) add the posted results-group labels and descriptions; the continuous \lm\ row is posted-record. $\IS = 1 - \mathrm{MSE}(\full)/\mathrm{MSE}(\blind)$; v2 values in \texttt{results/cr/main\_v2\_sensitivity.csv}.}
\label{tab:main}
\footnotesize
\setlength{\tabcolsep}{4pt}\resizebox{\linewidth}{!}{\begin{tabular}{llccc}
\toprule
Task & System & skill(\blind) & skill(\full) & interventional skill \\
\midrule
\multirow{8}{*}{Risk difference}
 & global mean & -- & 0 & -- \\
 & \gbt\ (5 seeds) & 0.297 \ci{0.220}{0.373} & 0.211 \ci{0.141}{0.281} & $-0.123$ \ci{-0.205}{-0.048} \\
 & \lm\ large, protocol-only & 0.462 \ci{0.338}{0.565} & 0.570 \ci{0.470}{0.655} & \textbf{0.201} \ci{0.083}{0.309} \\
 & \lm\ large, posted-record & 0.494 \ci{0.382}{0.587} & 0.629 \ci{0.547}{0.701} & \textbf{0.266} \ci{0.156}{0.373} \\
 & \lm\ small, protocol-only & 0.293 \ci{0.170}{0.394} & 0.351 \ci{0.238}{0.449} & \textbf{0.081} \ci{-0.013}{0.170} \\
 & \lm\ small, posted-record & 0.313 \ci{0.177}{0.425} & 0.389 \ci{0.268}{0.484} & \textbf{0.112} \ci{-0.012}{0.226} \\
 & Llama 3.1 70B, protocol-only & 0.356 \ci{0.270}{0.433} & 0.404 \ci{0.311}{0.483} & \textbf{0.075} \ci{-0.002}{0.151} \\
 & Llama 3.1 70B, posted-record & 0.360 \ci{0.271}{0.436} & 0.417 \ci{0.314}{0.499} & \textbf{0.089} \ci{0.018}{0.157} \\
\midrule
\multirow{2}{*}{Standardised difference} & \gbt\ (5 seeds) & 0.273 \ci{0.198}{0.348} & 0.288 \ci{0.220}{0.354} & $0.021$ \ci{-0.046}{0.082} \\
 & \lm\ large, posted-record & 0.259 \ci{0.125}{0.382} & 0.443 \ci{0.323}{0.560} & \textbf{0.249} \ci{0.069}{0.410} \\
\bottomrule
\end{tabular}}
\end{table}

%% file: _appendix1.tex
\appendix
\section{Benchmark construction and data card}
\label{app:construction}
\begin{table}[h]\centering\small
\caption{Cohort counts at each construction step (harvest of 9 September 2026; v2.1 pairing of 4 October 2026).}
\begin{tabular}{lr}\toprule Step & count \\ \midrule
Completed phase 2/3 interventional studies with posted results & 30{,}119 \\
Studies with a single-class numeric primary outcome and denominators & 23{,}755 \\
Arm records with $n\ge 20$ and finite outcome (binary / continuous) & 40{,}450 (18{,}548 / 21{,}902) \\
\quad trials contributing arm records & 18{,}077 \\
v2 contrasts (text-based pairing; after duplicate and $|d|>5$ removal) & 11{,}274 \\
\quad v2 trials (binary / continuous) & 8{,}011 (3{,}225 / 4{,}786) \\
v2.1: kept / re-paired / added / relabelled / dropped v2 contrasts & 9{,}383 / 375 / 240 / 6 / 1{,}510 \\
\quad removed by hand after the audit of new and re-paired test contrasts & 8 \\
v2.1 contrasts (binary / continuous) & 9{,}996 (4{,}014 / 5{,}982) \\
\quad trials contributing a contrast (binary / continuous) & 7{,}207 (2{,}950 / 4{,}257) \\
Contrasts by comparator: inert control / active comparator & 7{,}669 / 2{,}327 \\
Contrasts by split: train (posted $<$2023) / test ($\ge$2023) & 7{,}789 / 2{,}207 \\
Binary contrasts: train / test (trials) & 3{,}136 / 878 (655) \\
Continuous contrasts: train / test (trials) & 4{,}653 / 1{,}329 (978) \\
\bottomrule\end{tabular}\end{table}

\paragraph{Reproducibility and availability.} Harvest snapshot, benchmark tables (v2 and v2.1), data card, construction and pairing code, prompt templates, per-call language-model outputs, the pre-registration of the negative control and every result table are released at \url{\repourl} (MIT licence for code, CC~BY~4.0 for data and result tables)\doiclause{; the camera-ready release is archived with a DOI at \href{https://doi.org/\zenododoi}{doi:\zenododoi}}. The repository separates a deterministic layer from an API layer: \texttt{scripts/run\_camera\_ready.sh} regenerates every table in this paper from committed files with no network or model calls (one script per table under \texttt{scripts/cr/tables/}, writing \texttt{tables/<label>.tex} and a full-precision CSV), and \texttt{--check} re-fits every registry model and diffs the regenerated tables against the committed ones; the API layer (registry harvest and re-fetch, Claude and Llama calls, PubMedBERT embeddings) is documented with its archived outputs. Seeds are fixed; bootstrap intervals use 1{,}000 trial-level resamples (2{,}000 for the language-model arms), with one shared set of resamples for five-seed means. Figures are produced by \texttt{scripts/cr/figures/}.

\paragraph{Use of language models.} Language models appear in this work in four roles: as the systems under evaluation (Section~\ref{sec:lm}); as the judge of drug-class guesses (Section~\ref{sec:blind}); as the model-assisted reviewer of arm pairings in the audits, whose flagged cases were reviewed by the author; and as the labeller of abstracts in the workshop screen (Appendix~\ref{app:screen}), where 12 failed parses were labelled by hand. They were also used as a coding and writing assistant under the author's direction and review. Every number in the paper is produced by the released code from the released data and archived model outputs.

\paragraph{Pairing rules (v2.1).} The v2 pairing assigned arm roles from group text and paired the three largest treatment groups with the largest control of the preferred kind, with no check of molecule identity, design or stage. v2.1 adds, over registered arm types re-fetched for all 8{,}011 trials: (a)~\emph{same-molecule comparator}: a comparator whose normalised molecule key (components split on \emph{+, /, and, with, then}; placebo and standard-care components, doses, units and schedules removed) matches the treatment's is excluded and the next admissible comparator used, else the contrast is dropped (235 dropped, 180 re-paired); (b)~\emph{concurrency}: crossover and sequential designs are excluded, pooled and extension groups are excluded, and treatment and comparator must carry the same stage label (part, period, cohort, double-blind or open-label phase), staged groups being paired within stage rather than dropped; (c)~\emph{investigational arm}: where a trial has registered EXPERIMENTAL groups they form the treatment pool and the remaining control-role groups the comparator pool, with the v2 rule as fall-back. With all rules disabled the code reproduces v2 exactly. The universe is restricted to the 8{,}011 v2 trials; rule (c) would admit 1{,}516 further contrasts from 1{,}099 trials absent from v2, which are released separately and not audited. Arm types remain unknown for 7{,}690 result groups.

\paragraph{Audits.} Each audit samples contrasts and asks whether (i) the released values can be re-derived from the current registry record and (ii) the two arms form a valid concurrent treatment-versus-comparator pair of the stated kind; (ii) is model-assisted (Claude Sonnet~5 reading the record) with the same prompt for v2 and v2.1. v2: 100 sampled, 100 re-derived, 76 valid pairings; the 24 flagged cases were reviewed by the author (19 invalid, 5 valid, two of the latter with a wrong comparator-kind label only), and all 19 are absent from v2.1. v2.1: 100 sampled (seed 1), 100 re-derived, 96 valid (Wilson 95\% interval 0.90 to 0.98); the 4 failures are sub-group or cohort comparators and an interim randomisation, outside the three rules. All 125 new or re-paired test contrasts were audited individually: 115 valid; 7 pairing failures and 1 contrast whose comparator value cannot be re-derived from the current record were removed by hand, and 3 with a wrong comparator-kind label only were relabelled.

\paragraph{Intervention-field coverage.} After the corrected join (type-stripped intervention names, intervention-table arm-label back-references, fuzzy assignment of results-group titles to arm labels at threshold 0.55), 71.9\% of treatment arms are matched to a protocol arm and carry intervention names and types, 66.8\% carry an intervention description, 62.5\% a protocol arm description and 98.9\% a results-group description; 98.7\% of contrasts have at least 20 characters of intervention text across these fields. A preliminary build with a join defect (61\% names, 0\% descriptions) gave $\IS\approx 0$ for the language model too; the measurement is only as good as the completeness of the fields it withholds.

\paragraph{Arm-role rules.} \emph{Inert control}: registered arm type placebo comparator, sham comparator or no intervention, or title/description matching \emph{placebo, sham, vehicle, no treatment, no intervention, observation only, untreated, saline control} (placebo-labelled EXPERIMENTAL groups excepted under rule (c)). \emph{Active comparator}: arm type active comparator, or text matching \emph{active comparator/control, standard (of) care, usual care, reference drug/treatment/therapy, comparator drug/arm/therapy}. Otherwise \emph{treatment}.

\paragraph{Effect scales.} Binary: proportion $= $ count$/n_{\text{analysed}}$ (values reported as percentages divided by 100). Continuous: $d = (\bar y_T-\bar y_C)/s_p$ with $s_p^2 = [(n_T-1)s_T^2 + (n_C-1)s_C^2]/(n_T+n_C-2)$; $s$ from SD directly, from SE as $s = \mathrm{SE}\sqrt{n}$, or from a 95\% CI as $s=\sqrt{n}\,(u-l)/3.92$. Table~\ref{tab:design} reports the subsets with a participant-level binary unit and with arithmetic means only.

\paragraph{Therapeutic areas.} Regular expressions over the condition list, first match in order: oncology (\emph{cancer, carcinom, tumo(u)r, neoplas, leukemi, lymphom, myelom, melanom, sarcom, glioma}); cardiometabolic (\emph{diabet, cardiac, cardiovascular, heart, hypertens, cholesterol, lipid, obesity, coronary, stroke, atrial}); infectious; neurology/psychiatry; immune/inflammatory; other.

\paragraph{Noise ceiling.} For binary contrasts $\widehat{\mathrm{Var}}_{\text{samp}} = p_T(1-p_T)/n_T + p_C(1-p_C)/n_C$; for standardised differences $\widehat{\mathrm{Var}}_{\text{samp}} = (n_T+n_C)/(n_T n_C) + d^2/(2(n_T+n_C))$. Reliability $= 1 - \overline{\mathrm{Var}}_{\text{samp}}/\mathrm{Var}(\Delta)$: binary $0.906$ \ci{0.885}{0.922} on the v2.1 test split (mean sampling variance $0.0047$, total $0.0499$). The attainable interventional skill against a reference with error $\mathrm{MSE}_{\text{ref}}$ is $1 - \overline{\mathrm{Var}}_{\text{samp}}/\mathrm{MSE}_{\text{ref}}$: $0.87$ against the registry \blind\ model and $0.82$ against the large language model's blind prompt. No test contrast is noise-dominated.

\section{Features, models and hyperparameters}
\label{app:features}
\begin{table}[h]\centering\small
\caption{Feature groups. \blind\ = endpoint text + population + design + comparator type; \blind$^{-}$ omits comparator type; \full\ = \blind\ + intervention text + intervention type.}
\begin{tabular}{p{3.2cm}p{9.9cm}}\toprule Group & Fields \\ \midrule
Endpoint text & outcome title, outcome description, outcome timeframe, condition list \\
Intervention text & treatment arm: intervention names, intervention descriptions, protocol arm description, results-group title and description; control arm: intervention names, arm description, group title and description \\
Population/design (categorical) & phase, sponsor class, allocation, masking, primary purpose, therapeutic area, sex eligibility \\
Population (numeric) & $n_T$, $n_C$, mean age, fraction female, results-posting year \\
Comparator type & inert control vs active comparator \\
Intervention type & drug / biological / device / procedure / behavioural / \ldots (treatment arm) \\
\bottomrule\end{tabular}\end{table}
Text fields are concatenated, vectorised with TF-IDF (1--2-grams, English stop words, $\min\_df=3$, sublinear tf, 60{,}000 features) fitted on the training split, and reduced with truncated SVD to 120 components (random state 0). Categoricals are ordinal-encoded with unknowns mapped to $-1$; numerics are passed as-is with missing values left to the tree model. The regressor is scikit-learn's \texttt{HistGradientBoostingRegressor} (max\_iter 400, learning rate 0.06, min\_samples\_leaf 20, $\ell_2$ regularisation 1.0, early stopping on a 15\% validation fraction, random state = seed). Alternative architectures (Table~\ref{tab:arch}): separate 120-component SVD blocks for endpoint and intervention text; ridge on the sparse TF-IDF with $\alpha$ chosen by trial-grouped 5-fold cross-validation; a two-stage model that fits ridge on intervention text to the \blind\ model's out-of-fold residuals; and 768-dimensional sentence embeddings (S-PubMedBert-MS-MARCO) of each text block under gradient boosting and ridge. Same-molecule retrieval: molecule keys are normalised intervention names (type prefix and dose/route/salt tokens stripped; fallback to results-group title); matches are training contrasts sharing a key and the comparator type; the prediction is $\hat\Delta_{\blind} + \frac{n}{n+k}\,\bar r$ with $\bar r$ the mean out-of-fold blind residual of the matched contrasts and $k=3$. Variants keyed on intervention MeSH terms use terms re-fetched in October 2026 and are reported as provisional (Table~\ref{tab:molmesh}).

\section{Conformal and selective prediction}
\label{app:conformal}
\emph{Cross-fitted recipe} (main text): 5-fold GroupKFold over the training period supplies out-of-fold residuals for every training contrast and an out-of-fold error-scale model $\hat\sigma(x)$ (gradient boosting on $|r|$); the constant-width interval is $\hat\Delta \pm q$ with $q$ the $\lceil (n+1)(1-\alpha)\rceil/n$ empirical quantile of $|r|$ over the training pool, the normalised interval $\hat\Delta \pm q_\sigma\,\hat\sigma(x)$ with $q_\sigma$ the corresponding quantile of $|r|/\hat\sigma$. \emph{Split recipe} (Tables~\ref{tab:coverage} and~\ref{tab:opsplit}): the training pool is split 50/50 by trial into a fit set and a calibration set and the same quantities are computed on the calibration set; its error-scale model is fit to in-sample residuals, which are compressed towards zero, so $\hat\sigma$ sits at its floor value for about three quarters of evaluated contrasts and cannot rank them, which is why the main text uses the cross-fitted recipe. Regimes: in-distribution (trial-grouped holdout within the training period, $n=640$), temporal ($\ge 2023$, $n=878$), leave-one-area-out (train on the other five areas, both periods). Coverage intervals are trial-level bootstraps; a trial-level variant that calibrates on one contrast per trial is reported alongside. Selective prediction ranks test contrasts by $\hat\sigma(x)$ (registry model) or by the stated interval width (language model) and reports MAE among the most confident fraction. Operating points (Table~\ref{tab:op}) choose the answering threshold on calibration data as the largest answered fraction whose calibration MAE does not exceed the target and report the realised MAE when that threshold is applied unchanged to the evaluated set; the submitted version chose thresholds on the evaluated set, which meets any target by construction, and those rows are kept as ``oracle''.

\section{Language-model arm}
\label{app:llm}
\paragraph{Protocol.} One call per contrast, no system prompt beyond the provider default, \texttt{max\_tokens} = 1500, batched with concurrency 8; Claude models at the provider's default temperature, Llama~3.1 70B Instruct at temperature 0 with JSON output mode on a dedicated BF16 deployment (2 H200 GPUs, deleted after the run; a 20-prompt gate checked parse rate and prediction diversity first). Contrast identifiers \texttt{c0000}--\texttt{c1019} are the v2 binary test contrasts sorted by (NCT id, outcome title, treatment group); v2.1 contrasts not in v2 carry \texttt{c21\_}/\texttt{k21\_} identifiers. Parse: first JSON object in the response with finite \texttt{point}, \texttt{low}, \texttt{high} and $\texttt{low}\le\texttt{high}$, clipped to $[-1,1]$ ($[-5,5]$ for standardised differences). One retry with the suffix \emph{``IMPORTANT: output must be exactly one JSON object with keys point, low, high and nothing else.''}; then recorded as failure. Every contrast in every arm received one accepted prediction per condition; the negative-control run on v2 had 3 refusals, which are excluded. Model identifiers returned: \texttt{claude-sonnet-5}, \texttt{claude-haiku-4-5-20251001}, \texttt{accounts/fireworks/models/llama-v3p1-70b-instruct}. Documented cutoffs used for the exposure strata: Llama~3.1, December 2023 (boundary 1 January 2024); Haiku~4.5, training data to July 2025 (reliable knowledge to February 2025); Sonnet~5, January 2026. The continuous-endpoint prompt differs only in its header (defining the standardised mean difference, with $|d|$ of 0.2/0.5/0.8 given as small/medium/large).

\paragraph{Protocol-only and posted-record prompts.} The posted-record prompt (the submitted version's) renders, for each arm, the results-group title and description from the posted results module together with the registered intervention names and descriptions, the analysed $n$ and baseline summaries. The protocol-only prompt removes every field that exists only after read-out: results-group titles and descriptions, analysed arm sizes and baseline age and sex summaries; it keeps phase, conditions, eligibility criteria, comparator type, the registered intervention names and descriptions and the protocol arm descriptions, and the endpoint title, description, timeframe and unit as registered. Both share the same instruction header, redaction and parser. Outcome-vocabulary audit: 13.7\% of v2 posted-record contrasts have a results-group description matching a pattern such as \emph{responded, achieved, remission, \%}; protocol-only prompts contain no results-group text.

\paragraph{Perturbation prompts.} \emph{Arm swap}: the posted-record condition-A prompt with the treatment and control blocks exchanged and the control-type line adjusted. \emph{Negative control}: condition A with every treatment-arm intervention field replaced by a fixed unrelated marketed drug (loratadine for cardiometabolic trials, so that the substitute is never a plausible treatment, ezetimibe elsewhere) and the arm label and description redacted of the original agent; hypotheses H1--H3 and their criteria were written to the repository before any call (\texttt{results/llm/cr/negative\_control\_preregistration.md}). \emph{Blind repeat}: condition B re-run unchanged. \emph{Class guess}: Haiku~4.5 reads condition B and returns the most likely drug class, mechanism and an example agent as JSON; a second Haiku~4.5 call judges class, mechanism and exact-agent agreement with the true treatment, and, for the chance reference, with the treatment of a randomly permuted contrast (seed 0).

\paragraph{Redaction (condition B).} The treatment-arm and control-arm lines are replaced by fixed strings (\emph{``the experimental intervention [withheld]''}, \emph{``[withheld]''}, and for the control \emph{``the control condition (inert control / active comparator)''}). Every retained free-text field is passed through a redactor that removes (i) every token of length $\ge 4$ from the contrast's own intervention/arm fields that is capitalised, contains a digit, or carries a pharmacological suffix (\emph{-mab, -nib, -ciclib, -parib, -tinib, -prazole, -sartan, -floxacin, -cycline, -mycin, -cillin, -glutide, -gliptin, -glitazone, -setron, -statin, -afil, -vir, -cept, -pril, -olol, -dipine, \ldots}), excluding a list of common trial words, and (ii) any remaining suffix-matching word. \emph{Audit:} 0 of 1{,}020 protocol-only blind prompts and 4 of 1{,}020 posted-record blind prompts retain a token that also occurs in that contrast's intervention fields (\emph{arms, primary, scale, both}).

\paragraph{Template (condition A; condition B differs only in the arm block and redaction).}
\begin{footnotesize}\begin{verbatim}
You are simulating the results of a randomized clinical trial. You are given protocol
information for one trial and one binary primary endpoint (an event or responder
proportion). Predict the observed risk difference between the two arms.

Definition: risk difference = (proportion of participants in the TREATMENT arm with the
endpoint) minus (proportion in the CONTROL arm with the endpoint). Both proportions are on a
0-1 fraction scale, so the risk difference lies in [-1, 1]. A higher endpoint value is not
necessarily clinically better - predict the sign as measured, not as desirable.

Respond with STRICT JSON only, no prose, no markdown fences:
{"point": <number>, "low": <number>, "high": <number>}
where point is your best point estimate and [low, high] is a 90% central interval for the
observed risk difference.

--- TRIAL ---
Phase: {phase}
Condition(s): {conditions}
Eligibility criteria (truncated): {eligibility_text[:1500]}

Endpoint title: {outcome_title}
Endpoint description: {outcome_desc[:900]}
Endpoint timeframe: {outcome_timeframe}
Endpoint unit as reported: {unit}

Treatment arm size (n): {n_treat}
Control arm size (n): {n_ctrl}
Control type: {inert control (placebo / sham / no-treatment) |
               active comparator (another active treatment)}
Population mean age (years): {age_mean}
Population fraction female: {frac_female}

TREATMENT arm label: {treat_group}
TREATMENT arm intervention name(s): {treat_intervention}
TREATMENT arm intervention description: {treat_intervention_desc[:600]}
TREATMENT arm description: {treat_arm_desc[:600]} {treat_group_desc[:600]}
CONTROL arm label: {ctrl_group}
CONTROL arm intervention name(s): {ctrl_intervention}
CONTROL arm intervention description: {ctrl_intervention_desc[:600]}
CONTROL arm description: {ctrl_arm_desc[:600]} {ctrl_group_desc[:600]}

--- END TRIAL ---
Return only the JSON object.
\end{verbatim}\end{footnotesize}

\input{tables/lmruns}
\input{tables/contamination}
\input{tables/exposure}

%% file: tables/lmruns.tex
\begin{table}[h]\centering\footnotesize
\caption{All language-model runs. Upper block: benchmark arms on the v2.1 temporal test split; protocol-only prompts contain registered protocol fields only, posted-record prompts are the original prompts (continuous uses the posted-record prompt). Lower block: perturbation runs of Claude Sonnet~5 with the original (posted-record) prompts on the 878 v2.1 binary test contrasts (the 62 new or re-paired contrasts were run separately with identical prompts); the repeat run of condition~A exists for v2 only and is reported on the 816 v2.1 contrasts it covers (`v2.1 kept'). v2 values in \texttt{results/cr/lmruns.csv}. Skill against the family's training mean; $\IS$ = \full\ (A) vs \blind\ (B) with 95\% trial-level bootstrap intervals (2{,}000 resamples); coverage of the self-stated 90\% interval as A / B (MAE and mean width in the CSV). The negative-control row excludes 3 refusals; the arm-swap row is scored against the negated truth. Run names are abbreviated (full names in the CSV). Class guess (Claude Haiku~4.5 from the condition-B prompt, judged by Haiku~4.5; v2.1, 878 contrasts): the drug class matches the true treatment in 0.449 vs 0.023 for a permuted treatment (876 judged), mechanism 0.454 vs 0.019, exact agent 0.208 vs 0.007.}
\label{tab:lmruns}
\resizebox{\linewidth}{!}{\begin{tabular}{lllrrcccc}\toprule run & family & test set & $n$ & trials & skill(A) & skill(B) & $\IS$ & coverage A / B \\ \midrule
Sonnet 5, protocol & binary & v2.1 & 878 & 655 & 0.570 & 0.462 & $0.201$ \ci{0.083}{0.309} & 0.70 / 0.69 \\
Sonnet 5, posted & binary & v2.1 & 878 & 655 & 0.629 & 0.494 & $0.266$ \ci{0.156}{0.373} & 0.74 / 0.70 \\
Haiku 4.5, protocol & binary & v2.1 & 878 & 655 & 0.351 & 0.293 & $0.081$ \ci{-0.013}{0.170} & 0.58 / 0.59 \\
Haiku 4.5, posted & binary & v2.1 & 878 & 655 & 0.389 & 0.313 & $0.112$ \ci{-0.012}{0.226} & 0.57 / 0.61 \\
Llama 3.1 70B, protocol & binary & v2.1 & 878 & 655 & 0.404 & 0.356 & $0.075$ \ci{-0.002}{0.151} & 0.59 / 0.63 \\
Llama 3.1 70B, posted & binary & v2.1 & 878 & 655 & 0.417 & 0.360 & $0.089$ \ci{0.018}{0.157} & 0.66 / 0.69 \\
Sonnet 5, posted & continuous & v2.1 & 1{,}329 & 978 & 0.443 & 0.259 & $0.249$ \ci{0.069}{0.410} & 0.77 / 0.67 \\
\midrule
Sonnet 5, repeat run A vs original B & binary & v2.1 kept & 816 & 632 & 0.629 & 0.487 & $0.277$ \ci{0.169}{0.379} & 0.75 / 0.70 \\
Sonnet 5, original A vs repeat run B & binary & v2.1 & 878 & 655 & 0.629 & 0.482 & $0.283$ \ci{0.158}{0.396} & 0.74 / 0.71 \\
Sonnet 5, neg. control (drug substituted) vs original B & binary & v2.1 & 878 & 655 & 0.055 & 0.494 & $-0.869$ \ci{-1.335}{-0.475} & 0.66 / 0.70 \\
Sonnet 5, arms swapped (vs negated truth) & binary & v2.1 & 878 & 655 & 0.623 & -- & -- & 0.75 / -- \\
\bottomrule\end{tabular}}\end{table}

%% file: tables/contamination.tex
\begin{table}[h]\centering\small
\caption{Interventional skill inside and outside each model's uncontaminated set, v2.1 binary temporal test split (878 contrasts). A trial is uncontaminated for a model when its primary completion and first results posting both fall on or after the model's boundary date and no linked results publication predates it. Language-model columns use protocol-only and posted-record prompts (2{,}000 trial resamples); the registry \gbt\ column (five-seed mean, one shared set of 1{,}000 trial resamples) has no pretraining exposure and shows the effect of the date stratum alone. Strata with fewer than 30 contrasts ($^{\dagger}$) are not interpretable and are shown without intervals. For Sonnet~5 the training and reliable-knowledge boundaries coincide. Continuous-family strata are in \texttt{results/cr/contamination.csv}.}
\label{tab:contamination}
\resizebox{\linewidth}{!}{\begin{tabular}{lllrrccc}\toprule model & boundary & stratum & contrasts & trials & $\IS$ protocol-only & $\IS$ posted-record & $\IS$ \gbt \\ \midrule
Llama 3.1 70B & training cutoff (2024-01-01) & uncontaminated & 217 & 152 & $0.010$ \ci{-0.103}{0.123} & $0.089$ \ci{-0.012}{0.195} & $-0.058$ \ci{-0.214}{0.104} \\
 &  & complement & 661 & 503 & $0.097$ \ci{0.005}{0.188} & $0.089$ \ci{0.006}{0.169} & $-0.144$ \ci{-0.232}{-0.052} \\
\addlinespace
Haiku 4.5 & training cutoff (2025-08-01) & uncontaminated & 5 & 3 & $-0.102^{\dagger}$ & $0.579^{\dagger}$ & $-0.759^{\dagger}$ \\
 &  & complement & 873 & 652 & $0.083$ \ci{-0.016}{0.171} & $0.107$ \ci{-0.021}{0.221} & $-0.122$ \ci{-0.196}{-0.046} \\
\addlinespace
Haiku 4.5 & reliable-knowledge cutoff (2025-03-01) & uncontaminated & 52 & 33 & $0.060$ \ci{-0.256}{0.331} & $0.146$ \ci{-0.144}{0.337} & $0.008$ \ci{-0.208}{0.152} \\
 &  & complement & 826 & 622 & $0.082$ \ci{-0.019}{0.174} & $0.110$ \ci{-0.016}{0.221} & $-0.132$ \ci{-0.211}{-0.048} \\
\addlinespace
Sonnet 5 & training / reliable-knowledge cutoff (2026-02-01) & uncontaminated & 0 & 0 & -- & -- & -- \\
 &  & complement & 878 & 655 & $0.201$ \ci{0.083}{0.309} & $0.266$ \ci{0.156}{0.373} & $-0.123$ \ci{-0.205}{-0.048} \\
\bottomrule\end{tabular}}\end{table}

%% file: tables/exposure.tex
\begin{table}[h]\centering\footnotesize
\caption{Exposure test with Llama~3.1 70B (knowledge cutoff December 2023). Interventional skill with 95\% trial-level bootstrap
intervals (2{,}000 resamples). Uncontaminated: primary completion and first results posting on or after 2024-01-01 and no
earlier linked results publication. ``Molecule seen'': a normalised treatment name occurs among treatment arms in the registry
training split (a proxy for prior knowledge, not a measure of pretraining exposure). ``Flagged'': results-group descriptions
match an outcome-vocabulary pattern.}\label{tab:exposure}
\begin{tabular}{llccc}\toprule
set & stratum / split & prompt & $\IS$ & $n$ (trials) \\ \midrule
v2.1 & all & protocol-only & $0.08$ \ci{0.00}{0.15} & 878 (655) \\
v2.1 & all & posted-record & $0.09$ \ci{0.02}{0.16} & 878 (655) \\
v2.1 & uncontaminated & protocol-only & $0.01$ \ci{-0.10}{0.12} & 217 (152) \\
v2.1 & uncontaminated & posted-record & $0.09$ \ci{-0.01}{0.20} & 217 (152) \\
v2.1 & complement & protocol-only & $0.10$ \ci{0.01}{0.19} & 661 (503) \\
v2.1 & complement & posted-record & $0.09$ \ci{0.01}{0.17} & 661 (503) \\
v2.1 & uncontaminated, vocabulary unflagged & posted-record & $0.08$ \ci{-0.03}{0.19} & 196 (134) \\
v2.1 & uncontaminated, vocabulary flagged & posted-record & $0.16$ \ci{-0.05}{0.33} & 21 (18) \\
v2.1 & uncontaminated, molecule seen & protocol-only & $0.14$ \ci{-0.13}{0.34} & 58 (46) \\
v2.1 & uncontaminated, molecule unseen & protocol-only & $-0.01$ \ci{-0.15}{0.11} & 159 (107) \\
v2.1 & complement, vocabulary unflagged & posted-record & $0.09$ \ci{-0.01}{0.18} & 564 (421) \\
v2.1 & complement, vocabulary flagged & posted-record & $0.09$ \ci{-0.01}{0.20} & 97 (83) \\
v2.1 & complement, molecule seen & protocol-only & $0.16$ \ci{0.02}{0.29} & 242 (200) \\
v2.1 & complement, molecule unseen & protocol-only & $0.07$ \ci{-0.04}{0.19} & 419 (310) \\
v2 & all & protocol-only & $0.12$ \ci{0.04}{0.20} & 1020 (740) \\
v2 & all & posted-record & $0.12$ \ci{0.04}{0.19} & 1020 (740) \\
v2 & uncontaminated & protocol-only & $0.03$ \ci{-0.08}{0.12} & 245 (172) \\
v2 & uncontaminated & posted-record & $0.12$ \ci{0.03}{0.21} & 245 (172) \\
v2 & complement & protocol-only & $0.15$ \ci{0.05}{0.24} & 775 (568) \\
v2 & complement & posted-record & $0.12$ \ci{0.01}{0.21} & 775 (568) \\
v2 & uncontaminated, vocabulary unflagged & posted-record & $0.13$ \ci{0.03}{0.22} & 215 (149) \\
v2 & uncontaminated, vocabulary flagged & posted-record & $0.05$ \ci{-0.13}{0.22} & 30 (24) \\
v2 & uncontaminated, molecule seen & protocol-only & $0.21$ \ci{-0.04}{0.35} & 57 (46) \\
v2 & uncontaminated, molecule unseen & protocol-only & $0.00$ \ci{-0.11}{0.09} & 188 (127) \\
v2 & complement, vocabulary unflagged & posted-record & $0.12$ \ci{0.00}{0.22} & 665 (478) \\
v2 & complement, vocabulary flagged & posted-record & $0.08$ \ci{-0.03}{0.18} & 110 (92) \\
v2 & complement, molecule seen & protocol-only & $0.14$ \ci{-0.02}{0.28} & 250 (208) \\
v2 & complement, molecule unseen & protocol-only & $0.15$ \ci{0.02}{0.26} & 525 (371) \\
\bottomrule\end{tabular}\end{table}

%% file: _appendix2.tex
\clearpage
\section{Full result tables}
\label{app:tables}
All tables are regenerated by the driver script \texttt{run\_camera\_ready.sh}; v2 (submitted-version) values are kept in the released CSVs and cited where they differ materially. Tables~\ref{tab:fg} and~\ref{tab:molmesh} rest on definitions that had to be re-derived because the submitted version's probe code was not preserved; Table~\ref{tab:fg} is reported afresh on v2.1 and Table~\ref{tab:molmesh} is flagged as provisional.

\input{tables/seeds}
\input{tables/robust}
\input{tables/arch}
\input{tables/fg}
\input{tables/seen}
\input{tables/level}
\input{tables/mol}
\input{tables/molmesh}
\input{tables/design}
\input{tables/designcont}
\input{tables/lmrevision}
\input{tables/blindness}
\input{tables/coverage}
\input{tables/op}
\input{tables/opsplit}
\input{tables/lmarea}
\input{tables/lmsel}

\begin{figure}[h]\centering
\includegraphics[width=\linewidth]{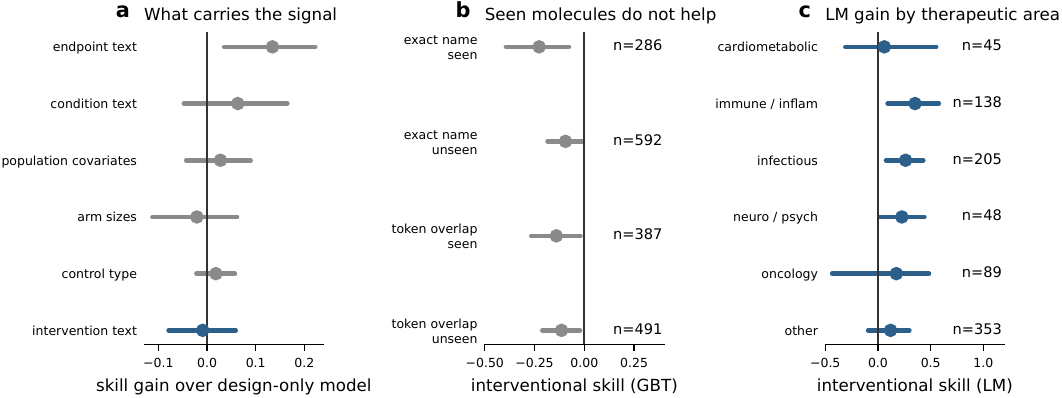}
\caption{\textbf{Probes (v2.1).} (a)~Feature-group gains over a design-only registry model (Table~\ref{tab:fg}). (b)~Interventional skill of the registry model for test contrasts whose intervention was or was not seen in training, two matching rules (Table~\ref{tab:seen}): seen molecules do not help. (c)~Large language model, protocol-only prompt, interventional skill by therapeutic area (Table~\ref{tab:lmarea}).}
\label{fig:probes}
\end{figure}

\clearpage
\section{Screen of NeurIPS 2025 workshop papers}
\label{app:screen}
\begin{figure}[h]\centering
\includegraphics[width=0.95\linewidth]{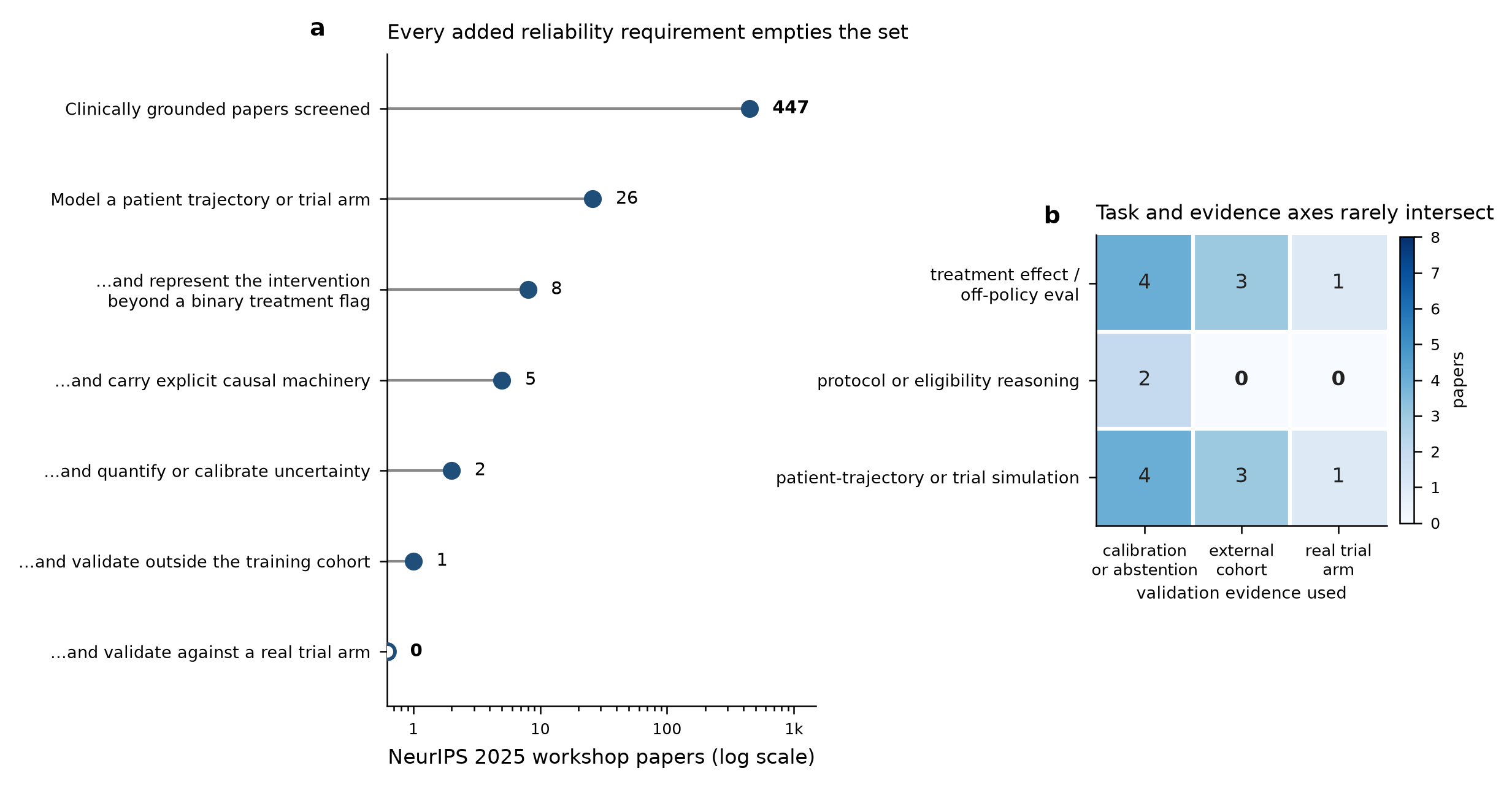}
\caption{\textbf{NeurIPS 2025 workshop screen.} (a)~Cumulative filter over 447 clinically grounded papers. (b)~Co-occurrence of task and validation-evidence axes.}
\label{fig:screen}
\end{figure}

To ground the claim that interventional validity is unmeasured in current practice we indexed the accepted papers of 19 NeurIPS 2025 workshops in health, causality, world models and reliability (1{,}648 papers; 1{,}618 abstracts retrieved), selected the 447 clinically grounded ones by workshop membership or clinical vocabulary, and labelled each abstract on seven axes (data modality, task, intervention representation, causal machinery, reliability mechanism, validation evidence, contribution type) with a language model, hand-labelling 12 that failed to parse. Of the 447, 26 model a patient trajectory or a trial arm; 8 of those represent the intervention beyond a binary flag; 5 add explicit causal machinery; 2 quantify uncertainty; 1 validates outside its training cohort; 0 validate against a real trial arm (Figure~\ref{fig:screen}a). The filter is cumulative, which is why its last row is 0 while Figure~\ref{fig:screen}b shows two papers in the real-trial-arm column: a tumour-dynamics survival model (AI4D3) that reports historical-trial replication but represents no intervention, and so leaves the filter at the second step, and a vasopressor-dosing reinforcement-learning paper (TS4H) scored against trial outcomes, which is a treatment-effect paper rather than a trajectory or arm simulator and never enters the filter. Neither both represents the intervention and validates against a real trial arm. Across all 447, target-trial emulation, conformal prediction and abstention each appear once; 233 validate on an internal split only. Labels are abstract-level and under-report validation detail; the index and labels are released with the code.

%% file: tables/seeds.tex
\begin{table}[h]\centering\small
\caption{Gradient boosting, binary risk difference, v2.1 temporal test split (878 contrasts / 655 trials), by seed. Columns 2--5: skill against the global training mean. \blind$^{-}$ omits comparator type; \blind\ includes it and is the reference for $\IS$. The mean-row interval applies one shared set of 1{,}000 trial resamples to all seeds. v2 values in the released CSV.}
\label{tab:seeds}
\begin{tabular}{lccccc}\toprule seed & \blind$^{-}$ & \blind & \full & intervention only & $\IS$ (\full\ vs \blind) \\ \midrule
0 & 0.261 & 0.295 & 0.216 & 0.030 & $-0.113$ \ci{-0.192}{-0.037} \\
1 & 0.250 & 0.301 & 0.210 & 0.069 & $-0.130$ \ci{-0.238}{-0.033} \\
2 & 0.261 & 0.311 & 0.209 & 0.050 & $-0.148$ \ci{-0.255}{-0.054} \\
3 & 0.249 & 0.283 & 0.208 & 0.043 & $-0.104$ \ci{-0.200}{-0.025} \\
4 & 0.262 & 0.295 & 0.212 & 0.078 & $-0.118$ \ci{-0.207}{-0.039} \\
\midrule mean & 0.257 & 0.297 & 0.211 & 0.054 & $-0.123$ \ci{-0.205}{-0.048} \\
\bottomrule\end{tabular}\end{table}

%% file: tables/robust.tex
\begin{table}[h]\centering\footnotesize
\caption{Robustness of the registry-model null on v2.1 (headline split: 878 contrasts / 655 trials): alternative temporal cuts (retrained), comparator/phase strata and the rich-intervention-text subset (headline models evaluated on the stratum, and retrained within it). Skill against the training mean; $\IS$ = skill(\full\,|\,\blind) with 95\% trial-level bootstrap intervals (1{,}000 resamples). v2 values in the released CSV.}
\label{tab:robust}
\resizebox{\linewidth}{!}{\begin{tabular}{llrrccc}\toprule setting & mode & contrasts & trials & skill(\blind) & skill(\full) & $\IS$ \\ \midrule
headline (seed 0) & headline & 878 & 655 & 0.295 & 0.216 & $-0.113$ \ci{-0.192}{-0.037} \\
cut=2021 & retrained & 1406 & 1047 & 0.273 & 0.245 & $-0.039$ \ci{-0.093}{0.019} \\
cut=2022 & retrained & 1102 & 821 & 0.263 & 0.219 & $-0.060$ \ci{-0.146}{0.015} \\
cut=2024 & retrained & 645 & 474 & 0.315 & 0.240 & $-0.109$ \ci{-0.248}{0.022} \\
ctrl\_kind=inert\_control & evaluated & 644 & 470 & 0.300 & 0.215 & $-0.121$ \ci{-0.209}{-0.028} \\
ctrl\_kind=inert\_control & retrained & 644 & 470 & 0.268 & 0.244 & $-0.032$ \ci{-0.127}{0.048} \\
ctrl\_kind=active\_comparator & evaluated & 234 & 185 & 0.269 & 0.222 & $-0.064$ \ci{-0.177}{0.046} \\
ctrl\_kind=active\_comparator & retrained & 234 & 185 & $-0.007$ & $-0.107$ & $-0.099$ \ci{-0.229}{0.015} \\
phase contains PHASE3 & evaluated & 445 & 383 & 0.280 & 0.233 & $-0.066$ \ci{-0.169}{0.033} \\
phase contains PHASE3 & retrained & 445 & 383 & 0.245 & 0.237 & $-0.011$ \ci{-0.115}{0.085} \\
phase PHASE2 only & evaluated & 366 & 234 & 0.335 & 0.216 & $-0.180$ \ci{-0.321}{-0.032} \\
phase PHASE2 only & retrained & 366 & 234 & 0.179 & 0.240 & 0.075 \ci{-0.047}{0.166} \\
rich intervention text (arm matched, desc non-empty) & evaluated & 697 & 528 & 0.293 & 0.227 & $-0.095$ \ci{-0.188}{-0.007} \\
rich intervention text (arm matched, desc non-empty) & retrained & 697 & 528 & 0.241 & 0.217 & $-0.032$ \ci{-0.126}{0.061} \\
\bottomrule\end{tabular}}\end{table}

%% file: tables/arch.tex
\begin{table}[h]\centering\footnotesize
\caption{Registry-model architectures on the v2.1 binary temporal test split (878 contrasts, 655 trials). Skill columns are against the global training mean; \blind\ includes comparator type, and $\IS$ is \full\ vs \blind\ of the same architecture. The headline GBT embeds endpoint and intervention text in one shared 120-dimensional SVD space; the other rows remove that sharing. Five-seed rows report the seed mean with an interval from one shared set of 1{,}000 trial resamples; ridge rows are deterministic, with $\alpha$ chosen by trial-grouped 5-fold cross-validation. The two-stage model adds a shrunken correction to the \blind\ prediction and so stays near zero by construction. No row has a positive interval. v2 values in the released CSV.}
\label{tab:arch}
\resizebox{\linewidth}{!}{\begin{tabular}{lcccc}\toprule architecture & seeds & skill(\blind) & skill(\full) & $\IS$ \\ \midrule
Headline GBT, shared SVD (120) & 5 & 0.30 & 0.21 & $$-0.12$$ \ci{$-0.20$}{$-0.05$} \\
Headline GBT, intervention text only & 5 & -- & 0.05 & -- \\
GBT, separate SVD blocks (120+120) & 5 & 0.31 & 0.28 & $$-0.04$$ \ci{$-0.08$}{$-0.01$} \\
Ridge on sparse TF-IDF, no SVD ($\alpha$ 1/1) & 1 & 0.27 & 0.24 & $$-0.04$$ \ci{$-0.10$}{0.02} \\
Ridge, intervention text only ($\alpha$ 1) & 1 & -- & 0.05 & -- \\
Two-stage: ridge blind + ridge(intervention) on OOF residuals ($\alpha$ 100) & 1 & 0.27 & 0.27 & $0.00$ \ci{0.00}{0.01} \\
GBT on PubMedBERT embeddings (768+768) & 5 & 0.29 & 0.29 & $$-0.01$$ \ci{$-0.04$}{0.02} \\
Ridge on PubMedBERT embeddings ($\alpha$ 0.3/1) & 1 & 0.28 & 0.27 & $$-0.01$$ \ci{$-0.05$}{0.03} \\
Ridge, PubMedBERT intervention embedding only ($\alpha$ 1) & 1 & -- & 0.07 & -- \\
\bottomrule\end{tabular}}\end{table}

%% file: tables/fg.tex
\begin{table}[h]\centering\small
\caption{Feature-group gains (gradient boosting, seed 0, v2.1 binary test split, 878 contrasts / 655 trials). ``design-only'' = phase, sponsor class, allocation, masking, purpose, area, sex. 95\% trial-level bootstrap intervals (1{,}000 resamples). Group definitions follow \texttt{ivbench.specs}; this table is computed afresh on v2.1 and supersedes the submitted version, whose probe code was not preserved.}
\label{tab:fg}
\begin{tabular}{llc}\toprule model & reference & skill \\ \midrule
design-only & train\_mean & 0.069 \ci{0.018}{0.118} \\
design $+$ endpoint text & design-only & 0.134 \ci{0.035}{0.222} \\
design $+$ endpoint text & train\_mean & 0.195 \ci{0.110}{0.280} \\
design $+$ condition text & design-only & 0.063 \ci{-0.047}{0.165} \\
design $+$ condition text & train\_mean & 0.128 \ci{0.026}{0.216} \\
design $+$ population covariates & design-only & 0.028 \ci{-0.043}{0.090} \\
design $+$ population covariates & train\_mean & 0.095 \ci{0.032}{0.159} \\
design $+$ arm sizes & design-only & $-0.021$ \ci{-0.112}{0.062} \\
design $+$ arm sizes & train\_mean & 0.050 \ci{-0.039}{0.124} \\
design $+$ control type & design-only & 0.018 \ci{-0.022}{0.057} \\
design $+$ control type & train\_mean & 0.086 \ci{0.032}{0.139} \\
design $+$ intervention text & design-only & $-0.009$ \ci{-0.079}{0.059} \\
design $+$ intervention text & train\_mean & 0.061 \ci{0.002}{0.113} \\
confound probe: sponsor\_class$+$phase$+$year & train\_mean & 0.008 \ci{-0.023}{0.041} \\
\bottomrule\end{tabular}\end{table}

%% file: tables/seen.tex
\begin{table}[h]\centering\small
\caption{Seen vs unseen interventions on v2.1 (878 binary test contrasts; five-seed means; headline models evaluated on the stratum). ``Seen'' = the test contrast's intervention key appears among training contrasts. Intervals: one shared set of 1{,}000 trial resamples applied to all seeds. Seen molecules do not help: $\IS$ is more negative where the molecule was in training. v2 values in the released CSV.}
\label{tab:seen}
\resizebox{\linewidth}{!}{\begin{tabular}{llrrccc}\toprule matching & stratum & contrasts & trials & skill(\blind) & skill(\full) & $\IS$ \\ \midrule
exact normalized name & seen & 286 & 230 & 0.425 & 0.296 & $-0.227$ \ci{-0.394}{-0.077} \\
exact normalized name & unseen & 592 & 435 & 0.251 & 0.180 & $-0.095$ \ci{-0.185}{-0.007} \\
token-Jaccard$\geq$0.5 & seen & 387 & 308 & 0.375 & 0.288 & $-0.140$ \ci{-0.267}{-0.019} \\
token-Jaccard$\geq$0.5 & unseen & 491 & 356 & 0.252 & 0.166 & $-0.114$ \ci{-0.211}{-0.022} \\
\bottomrule\end{tabular}}\end{table}

%% file: tables/level.tex
\begin{table}[h]\centering\small
\caption{Arm-level task on v2.1 (predicting each test arm's event proportion; 4{,}177 arms / 1{,}945 trials; pooled or total groups excluded, repaired arm roles). ``blind\_norole'' additionally omits the arm's role. 95\% trial-level bootstrap intervals (1{,}000 resamples). v2 values in the released CSV.}
\label{tab:level}
\begin{tabular}{llc}\toprule model & reference & skill \\ \midrule
blind & train\_mean & 0.245 \ci{0.204}{0.287} \\
blind\_norole & train\_mean & 0.225 \ci{0.187}{0.266} \\
full & train\_mean & 0.212 \ci{0.168}{0.255} \\
full & blind & $-0.044$ \ci{-0.090}{-0.003} \\
full & blind\_norole & $-0.018$ \ci{-0.062}{0.024} \\
\bottomrule\end{tabular}\end{table}

%% file: tables/mol.tex
\begin{table}[h]\centering\footnotesize
\caption{Same-molecule retrieval positive control on v2.1 (binary test split, $k=3$ matched training contrasts of the same normalised treatment name). ``residual'' adds the shrunk mean out-of-fold \blind\ residual of the matched contrasts to the \blind\ prediction; ``level'' replaces the prediction by their mean delta shrunk towards the training mean. sameCtrl restricts matches to the same comparator kind. 95\% trial-level bootstrap intervals (1{,}000 resamples). Last column: the gradient-boosting \full\ model's $\IS$ on the same subset. Variants keyed on intervention MeSH terms are in Table~\ref{tab:molmesh}.}
\label{tab:mol}
\resizebox{\linewidth}{!}{\begin{tabular}{lrrcc}\toprule variant & contrasts & trials & retrieval $\IS$ & \gbt\ $\IS$, same subset \\ \midrule
name\_sameCtrl\_k3\_residual & 246 & 201 & $-0.043$ \ci{-0.166}{0.052} & $-0.216$ \ci{-0.417}{-0.020} \\
name\_sameCtrl\_k3\_level & 246 & 201 & $-0.689$ \ci{-1.105}{-0.385} & $-0.216$ \ci{-0.417}{-0.020} \\
name\_anyCtrl\_k3\_residual & 286 & 230 & $-0.088$ \ci{-0.296}{0.071} & $-0.255$ \ci{-0.440}{-0.090} \\
name\_anyCtrl\_k3\_level & 286 & 230 & $-0.747$ \ci{-1.200}{-0.428} & $-0.255$ \ci{-0.440}{-0.090} \\
\bottomrule\end{tabular}}\end{table}

%% file: tables/molmesh.tex
\begin{table}[h]\centering\footnotesize
\caption{Same-molecule retrieval, MeSH-keyed variants (provisional). union keys = normalised treatment names $\cup$ intervention MeSH terms; mesh keys = MeSH terms only. The harvest-time MeSH terms were not preserved; the terms used here were re-fetched from ClinicalTrials.gov in October 2026, so the matched subsets differ from those of the submitted version (on v2: 342 vs 358 contrasts for the union variant) and these rows should be read as indicative. Same conventions as Table~\ref{tab:mol}.}
\label{tab:molmesh}
\resizebox{\linewidth}{!}{\begin{tabular}{lrrcc}\toprule variant & contrasts & trials & retrieval $\IS$ & \gbt\ $\IS$, same subset \\ \midrule
primary\_union\_sameCtrl\_k3\_residual & 297 & 240 & $-0.040$ \ci{-0.153}{0.054} & $-0.198$ \ci{-0.384}{-0.022} \\
primary\_union\_sameCtrl\_k3\_level & 297 & 240 & $-0.614$ \ci{-0.940}{-0.358} & $-0.198$ \ci{-0.384}{-0.022} \\
union\_anyCtrl\_k3\_residual & 330 & 265 & $-0.089$ \ci{-0.258}{0.043} & $-0.202$ \ci{-0.386}{-0.043} \\
union\_anyCtrl\_k3\_level & 330 & 265 & $-0.686$ \ci{-1.093}{-0.406} & $-0.202$ \ci{-0.386}{-0.043} \\
mesh\_sameCtrl\_k3\_residual & 193 & 165 & $-0.040$ \ci{-0.163}{0.087} & $-0.093$ \ci{-0.349}{0.096} \\
mesh\_sameCtrl\_k3\_level & 193 & 165 & $-0.443$ \ci{-0.809}{-0.164} & $-0.093$ \ci{-0.349}{0.096} \\
mesh\_anyCtrl\_k3\_residual & 213 & 183 & $-0.139$ \ci{-0.393}{0.073} & $-0.119$ \ci{-0.353}{0.085} \\
mesh\_anyCtrl\_k3\_level & 213 & 183 & $-0.468$ \ci{-0.895}{-0.168} & $-0.119$ \ci{-0.353}{0.085} \\
\bottomrule\end{tabular}}\end{table}

%% file: tables/design.tex
\begin{table}[h]\centering\footnotesize
\caption{Interventional skill on design subsets of the v2.1 binary temporal test split. Randomized parallel-group: allocation and intervention model from an October 2026 re-fetch of the registry records. Participant-level unit: the binary outcome is a count or percentage of participants. \gbt: five-seed mean with one shared set of 1{,}000 trial resamples, seed SD across the five seeds; language models: single runs, 2{,}000 trial resamples. Haiku and Llama posted-record rows, seed-0 registry values and the noise ceiling per subset are in \texttt{results/cr/design.csv} and \texttt{results/cr/design\_noise\_ceiling.csv}.}
\label{tab:design}
\begin{tabular}{lcc}\toprule system & $\IS$ & seed SD \\ \midrule
\multicolumn{3}{l}{\emph{all test}: 878 contrasts, 655 trials} \\
\quad \gbt & $-0.123$ \ci{-0.205}{-0.048} & 0.017 \\
\quad Sonnet 5, protocol-only & $0.201$ \ci{0.083}{0.309} & -- \\
\quad Sonnet 5, posted-record & $0.266$ \ci{0.156}{0.373} & -- \\
\quad Haiku 4.5, protocol-only & $0.081$ \ci{-0.013}{0.170} & -- \\
\quad Llama 3.1 70B, protocol-only & $0.075$ \ci{-0.002}{0.151} & -- \\
\midrule
\multicolumn{3}{l}{\emph{randomized parallel-group}: 842 contrasts, 627 trials} \\
\quad \gbt & $-0.132$ \ci{-0.214}{-0.052} & 0.013 \\
\quad Sonnet 5, protocol-only & $0.216$ \ci{0.100}{0.332} & -- \\
\quad Sonnet 5, posted-record & $0.275$ \ci{0.165}{0.379} & -- \\
\quad Haiku 4.5, protocol-only & $0.076$ \ci{-0.025}{0.165} & -- \\
\quad Llama 3.1 70B, protocol-only & $0.083$ \ci{0.001}{0.162} & -- \\
\midrule
\multicolumn{3}{l}{\emph{not randomized parallel-group}: 36 contrasts, 28 trials} \\
\quad \gbt & $0.052$ \ci{-0.264}{0.283} & 0.149 \\
\quad Sonnet 5, protocol-only & $-0.110$ \ci{-0.691}{0.163} & -- \\
\quad Sonnet 5, posted-record & $0.131$ \ci{-0.121}{0.416} & -- \\
\quad Haiku 4.5, protocol-only & $0.176$ \ci{-0.208}{0.465} & -- \\
\quad Llama 3.1 70B, protocol-only & $-0.080$ \ci{-0.641}{0.180} & -- \\
\midrule
\multicolumn{3}{l}{\emph{participant-level unit}: 845 contrasts, 634 trials} \\
\quad \gbt & $-0.109$ \ci{-0.183}{-0.035} & 0.017 \\
\quad Sonnet 5, protocol-only & $0.186$ \ci{0.068}{0.295} & -- \\
\quad Sonnet 5, posted-record & $0.261$ \ci{0.147}{0.370} & -- \\
\quad Haiku 4.5, protocol-only & $0.072$ \ci{-0.025}{0.156} & -- \\
\quad Llama 3.1 70B, protocol-only & $0.087$ \ci{0.014}{0.164} & -- \\
\midrule
\multicolumn{3}{l}{\emph{participant-level unit \& randomized}: 821 contrasts, 614 trials} \\
\quad \gbt & $-0.114$ \ci{-0.200}{-0.031} & 0.016 \\
\quad Sonnet 5, protocol-only & $0.197$ \ci{0.079}{0.307} & -- \\
\quad Sonnet 5, posted-record & $0.262$ \ci{0.145}{0.374} & -- \\
\quad Haiku 4.5, protocol-only & $0.060$ \ci{-0.035}{0.148} & -- \\
\quad Llama 3.1 70B, protocol-only & $0.093$ \ci{0.020}{0.168} & -- \\
\midrule
\multicolumn{3}{l}{\emph{participant-level unit \& randomized parallel}: 809 contrasts, 606 trials} \\
\quad \gbt & $-0.119$ \ci{-0.199}{-0.034} & 0.016 \\
\quad Sonnet 5, protocol-only & $0.202$ \ci{0.077}{0.316} & -- \\
\quad Sonnet 5, posted-record & $0.269$ \ci{0.143}{0.383} & -- \\
\quad Haiku 4.5, protocol-only & $0.066$ \ci{-0.027}{0.151} & -- \\
\quad Llama 3.1 70B, protocol-only & $0.096$ \ci{0.022}{0.174} & -- \\
\bottomrule\end{tabular}\end{table}

%% file: tables/designcont.tex
\begin{table}[h]\centering\footnotesize
\caption{Interventional skill on design and valid-conversion subsets of the v2.1 continuous temporal test split (posted-record prompt for the language model). Strict: arithmetic means only; sensitivity rows also admit least-squares means. \gbt: five-seed mean with one shared set of 1{,}000 trial resamples, seed SD across the five seeds; language models: single runs, 2{,}000 trial resamples. Haiku and Llama posted-record rows, seed-0 registry values and the noise ceiling per subset are in \texttt{results/cr/design.csv} and \texttt{results/cr/design\_noise\_ceiling.csv}.}
\label{tab:designcont}
\begin{tabular}{lcc}\toprule system & $\IS$ & seed SD \\ \midrule
\multicolumn{3}{l}{\emph{all test}: 1{,}329 contrasts, 978 trials} \\
\quad \gbt & $0.021$ \ci{-0.046}{0.082} & 0.019 \\
\quad Sonnet 5, posted-record & $0.249$ \ci{0.069}{0.410} & -- \\
\midrule
\multicolumn{3}{l}{\emph{randomized parallel-group}: 1{,}305 contrasts, 961 trials} \\
\quad \gbt & $0.022$ \ci{-0.050}{0.081} & 0.019 \\
\quad Sonnet 5, posted-record & $0.245$ \ci{0.046}{0.404} & -- \\
\midrule
\multicolumn{3}{l}{\emph{not randomized parallel-group}: 24 contrasts, 17 trials} \\
\quad \gbt & $-0.009$ \ci{-0.337}{0.164} & 0.065 \\
\quad Sonnet 5, posted-record & $0.553$ \ci{0.306}{0.762} & -- \\
\midrule
\multicolumn{3}{l}{\emph{strict: arithmetic MEAN}: 484 contrasts, 374 trials} \\
\quad \gbt & $-0.030$ \ci{-0.129}{0.041} & 0.026 \\
\quad Sonnet 5, posted-record & $0.193$ \ci{-0.089}{0.432} & -- \\
\midrule
\multicolumn{3}{l}{\emph{strict \& randomized}: 474 contrasts, 366 trials} \\
\quad \gbt & $-0.031$ \ci{-0.125}{0.035} & 0.029 \\
\quad Sonnet 5, posted-record & $0.186$ \ci{-0.094}{0.432} & -- \\
\midrule
\multicolumn{3}{l}{\emph{strict \& randomized parallel}: 468 contrasts, 362 trials} \\
\quad \gbt & $-0.029$ \ci{-0.127}{0.042} & 0.028 \\
\quad Sonnet 5, posted-record & $0.185$ \ci{-0.089}{0.425} & -- \\
\midrule
\multicolumn{3}{l}{\emph{sensitivity: MEAN or LS MEAN \& randomized}: 1{,}113 contrasts, 789 trials} \\
\quad \gbt & $0.019$ \ci{-0.052}{0.083} & 0.025 \\
\quad Sonnet 5, posted-record & $0.251$ \ci{0.059}{0.418} & -- \\
\midrule
\multicolumn{3}{l}{\emph{sensitivity: MEAN or LS MEAN \& randomized parallel}: 1{,}102 contrasts, 782 trials} \\
\quad \gbt & $0.020$ \ci{-0.060}{0.085} & 0.025 \\
\quad Sonnet 5, posted-record & $0.249$ \ci{0.046}{0.422} & -- \\
\bottomrule\end{tabular}\end{table}

%% file: tables/lmrevision.tex
\begin{table}[h]\centering\footnotesize
\caption{Language-model revision checks. (a) Interventional skill with protocol-only and posted-record prompts on the v2.1 binary test split, with the paired difference on one shared set of 2{,}000 trial resamples. (b)--(d) use Claude Sonnet~5 with the original (posted-record) prompts on the 878 v2.1 binary test contrasts (v2 and 816-contrast values in \texttt{results/cr/lmrevision.csv}). (b) Condition A with the arms exchanged; A is the original condition-A prediction. (c) A second run of condition B. (d) Hypotheses H1--H3 exactly as pre-registered before any negative-control call (\texttt{results/llm/cr/negative\_control\_preregistration.md}): the treatment is replaced by loratadine (cardiometabolic trials) or ezetimibe (all other areas); NC is the substituted prediction; refused calls (3 on v2) are excluded. Intervals are 95\% trial-level bootstrap (2{,}000 resamples).}
\label{tab:lmrevision}
\resizebox{\linewidth}{!}{\begin{tabular}{lcrl}\toprule statistic & value & $n$ & pre-registered criterion / outcome \\ \midrule
\multicolumn{4}{l}{\emph{(a) prompt condition, v2.1}} \\
\quad Claude Sonnet 5: IS, protocol-only & $0.201$ \ci{0.083}{0.309} & 878 &  \\
\quad Claude Sonnet 5: IS, posted-record & $0.266$ \ci{0.156}{0.373} & 878 &  \\
\quad Claude Sonnet 5: IS difference, posted-record minus protocol-only & $0.066$ \ci{-0.035}{0.166} & 878 &  \\
\quad Claude Haiku 4.5: IS, protocol-only & $0.081$ \ci{-0.013}{0.170} & 878 &  \\
\quad Claude Haiku 4.5: IS, posted-record & $0.112$ \ci{-0.012}{0.226} & 878 &  \\
\quad Claude Haiku 4.5: IS difference, posted-record minus protocol-only & $0.030$ \ci{-0.090}{0.151} & 878 &  \\
\quad Llama 3.1 70B: IS, protocol-only & $0.075$ \ci{-0.002}{0.151} & 878 &  \\
\quad Llama 3.1 70B: IS, posted-record & $0.089$ \ci{0.018}{0.157} & 878 &  \\
\quad Llama 3.1 70B: IS difference, posted-record minus protocol-only & $0.014$ \ci{-0.067}{0.088} & 878 &  \\
\midrule
\multicolumn{4}{l}{\emph{(b) arm swap, v2.1}} \\
\quad sign reversal, $|A|$ $>$ 0.01 & $0.861$ \ci{0.830}{0.890} & 742 & a model that reads the arms returns about $-A$ \\
\quad corr(swapped, $-A$) & $0.947$ \ci{0.921}{0.967} & 878 &  \\
\quad mean(swapped + A) (ideal 0) & $0.008$ \ci{0.003}{0.014} & 878 &  \\
\quad mean $|A|$ (scale reference) & $0.152$ & 878 &  \\
\midrule
\multicolumn{4}{l}{\emph{(c) blind repeat, v2.1}} \\
\quad IS, original A vs original B & $0.266$ \ci{0.156}{0.373} & 878 &  \\
\quad IS, original A vs repeat B & $0.283$ \ci{0.158}{0.396} & 878 &  \\
\quad IS difference, repeat B minus original B & $0.017$ \ci{-0.024}{0.052} & 878 &  \\
\quad corr(repeat B, original B) & $0.981$ & 878 &  \\
\quad fraction of identical B points & $0.556$ & 878 &  \\
\midrule
\multicolumn{4}{l}{\emph{(d) pre-registered negative control, v2.1}} \\
\quad H1: mean($|A|$ - $|\mathrm{NC}|$) & $0.035$ \ci{0.020}{0.049} & 878 & $>$ 0, CI excludes 0: supported \\
\quad H1: fraction $|\mathrm{NC}|$ $<$ $|A|$ & $0.559$ \ci{0.519}{0.597} & 878 & $>$ 0.5: supported \\
\quad H1: fraction $|\mathrm{NC}|$ = $|A|$ (ties) & $0.151$ & 878 &  \\
\quad H1 overall & -- & 878 & both H1 conditions: supported \\
\quad H2 (inert control): mean $|A|$ & $0.186$ & 644 &  \\
\quad H2 (inert control): mean $|\mathrm{NC}|$ & $0.113$ & 644 &  \\
\quad H2 (inert control): mean $|A|$ - mean $|\mathrm{NC}|$ & $0.073$ \ci{0.057}{0.089} & 644 & $>$ 0: supported \\
\quad H2 (inert control): slope of NC on A through origin & $0.590$ \ci{0.498}{0.686} & 644 & $<$ 0.5: inconclusive \\
\quad H2 overall & -- & 644 & both H2 conditions: not supported \\
\quad H3 (active comparator, A $>$ 0): mean(NC - A) & $-0.118$ \ci{-0.178}{-0.068} & 113 & $<$ 0: supported \\
\quad secondary: skill(NC) vs training mean & $0.055$ \ci{-0.164}{0.247} & 878 &  \\
\quad secondary: skill(A $\mid$ NC) & $0.607$ \ci{0.500}{0.694} & 878 &  \\
\quad secondary: skill(NC $\mid$ B) & $-0.869$ \ci{-1.335}{-0.475} & 878 &  \\
\quad secondary: corr(NC, A) & $0.649$ & 878 &  \\
\quad secondary: corr(NC, B) & $0.583$ & 878 &  \\
\bottomrule\end{tabular}}\end{table}

%% file: tables/blindness.tex
\begin{table}[h]\centering\footnotesize
\caption{How blind is \blind? (a) Claude Haiku~4.5 reads the original condition-B prompt and guesses the withheld treatment's drug class, mechanism and an example agent; a Haiku~4.5 judge scores the guess against the true treatment and, as a chance reference, against the treatment of a randomly permuted contrast (judge calls that failed to parse are excluded). On the 878 v2.1 binary test contrasts (v2 values in \texttt{results/cr/blindness.csv}); 95\% trial-level bootstrap intervals (2{,}000 resamples). (b) For v2.1 test contrasts whose normalised treatment name occurs in the training split, the share for which a training contrast with the same molecule is among the $k$ nearest by \blind\ text (endpoint and condition fields), as TF-IDF and as the 120-dimensional SVD the registry model uses; chance is the hypergeometric rate for $k$ random training contrasts; 1{,}000 trial resamples.}
\label{tab:blindness}
\resizebox{\linewidth}{!}{\begin{tabular}{lcccr}\toprule statistic & observed & chance & ratio & $n$ (trials) \\ \midrule
\multicolumn{5}{l}{\emph{(a) language model: treatment class guessed from the \blind\ prompt (Claude Haiku~4.5, v2.1 binary)}} \\
\quad drug class & 0.449 \ci{0.408}{0.489} & 0.023 \ci{0.014}{0.033} & 19.7 & 878 (655) \\
\quad mechanism & 0.454 \ci{0.414}{0.494} & 0.019 \ci{0.011}{0.029} & 23.4 & 878 (655) \\
\quad exact agent & 0.208 \ci{0.179}{0.240} & 0.007 \ci{0.002}{0.013} & 30.4 & 878 (655) \\
\midrule
\multicolumn{5}{l}{\emph{(b) registry \blind\ text: same-molecule training contrast retrieved (v2.1)}} \\
\quad binary, tfidf, top-1 & 0.232 \ci{0.172}{0.298} & 0.003 & 82.2 & 228 (188) \\
\quad binary, tfidf, top-5 & 0.351 \ci{0.279}{0.429} & 0.014 & 25.1 & 228 (188) \\
\quad binary, tfidf, top-10 & 0.439 \ci{0.366}{0.518} & 0.027 & 16.0 & 228 (188) \\
\quad binary, svd120 (model input), top-1 & 0.193 \ci{0.136}{0.254} & 0.003 & 68.2 & 228 (188) \\
\quad binary, svd120 (model input), top-5 & 0.298 \ci{0.232}{0.368} & 0.014 & 21.4 & 228 (188) \\
\quad binary, svd120 (model input), top-10 & 0.386 \ci{0.314}{0.462} & 0.027 & 14.1 & 228 (188) \\
\quad continuous, tfidf, top-1 & 0.260 \ci{0.214}{0.309} & 0.003 & 75.1 & 516 (398) \\
\quad continuous, tfidf, top-5 & 0.370 \ci{0.321}{0.426} & 0.017 & 21.9 & 516 (398) \\
\quad continuous, tfidf, top-10 & 0.465 \ci{0.413}{0.523} & 0.033 & 14.1 & 516 (398) \\
\quad continuous, svd120 (model input), top-1 & 0.207 \ci{0.165}{0.255} & 0.003 & 60.0 & 516 (398) \\
\quad continuous, svd120 (model input), top-5 & 0.320 \ci{0.270}{0.375} & 0.017 & 18.9 & 516 (398) \\
\quad continuous, svd120 (model input), top-10 & 0.417 \ci{0.363}{0.471} & 0.033 & 12.6 & 516 (398) \\
\bottomrule\end{tabular}}\end{table}

%% file: tables/coverage.tex
\begin{table}[h]\centering\footnotesize
\caption{Conformal coverage and mean width (nominal 90\%) for the registry model, v2.1 binary contrasts, both calibration recipes, all regimes. Coverage intervals are 95\% trial-level bootstrap intervals (1{,}000 resamples over trials), not Wilson intervals. The last column recalibrates on one randomly chosen contrast per calibration trial, which restores trial-level exchangeability (mean over 200 draws, with the 2.5th--97.5th percentiles across draws); coverage of one random contrast per test trial is in \texttt{results/cr/coverage.csv}.}
\label{tab:coverage}
\resizebox{\linewidth}{!}{\begin{tabular}{lllrcccc}\toprule regime & recipe & interval & $n$ & coverage & width & MAE & coverage, one contrast per calibration trial \\ \midrule
in\_distribution & split & constant & 640 & 0.889 \ci{0.860}{0.919} & 0.626 & 0.138 & 0.880 \ci{0.875}{0.889} \\
in\_distribution & split & normalized & 640 & 0.845 \ci{0.813}{0.878} & 0.528 & 0.138 & 0.858 \ci{0.852}{0.866} \\
in\_distribution & crossfit & constant & 640 & 0.906 \ci{0.879}{0.933} & 0.598 & 0.124 & 0.893 \ci{0.891}{0.895} \\
in\_distribution & crossfit & normalized & 640 & 0.867 \ci{0.834}{0.897} & 0.514 & 0.124 & 0.868 \ci{0.866}{0.870} \\
temporal\_2023plus & split & constant & 878 & 0.888 \ci{0.861}{0.914} & 0.602 & 0.136 & 0.875 \ci{0.868}{0.880} \\
temporal\_2023plus & split & normalized & 878 & 0.867 \ci{0.839}{0.896} & 0.551 & 0.136 & 0.854 \ci{0.847}{0.862} \\
temporal\_2023plus & crossfit & constant & 878 & 0.874 \ci{0.847}{0.899} & 0.563 & 0.130 & 0.865 \ci{0.859}{0.867} \\
temporal\_2023plus & crossfit & normalized & 878 & 0.845 \ci{0.814}{0.873} & 0.501 & 0.130 & 0.844 \ci{0.841}{0.847} \\
loao\_cardiometabolic & split & constant & 254 & 0.949 \ci{0.917}{0.977} & 0.603 & 0.085 & 0.949 \ci{0.949}{0.949} \\
loao\_cardiometabolic & split & normalized & 254 & 0.921 \ci{0.880}{0.958} & 0.455 & 0.085 & 0.919 \ci{0.913}{0.921} \\
loao\_cardiometabolic & crossfit & constant & 254 & 0.949 \ci{0.914}{0.976} & 0.577 & 0.085 & 0.944 \ci{0.937}{0.945} \\
loao\_cardiometabolic & crossfit & normalized & 254 & 0.913 \ci{0.870}{0.951} & 0.388 & 0.085 & 0.914 \ci{0.913}{0.917} \\
loao\_immune\_inflam & split & constant & 874 & 0.857 \ci{0.825}{0.887} & 0.580 & 0.156 & 0.846 \ci{0.843}{0.848} \\
loao\_immune\_inflam & split & normalized & 874 & 0.810 \ci{0.778}{0.844} & 0.532 & 0.156 & 0.806 \ci{0.803}{0.811} \\
loao\_immune\_inflam & crossfit & constant & 874 & 0.843 \ci{0.810}{0.875} & 0.563 & 0.158 & 0.830 \ci{0.824}{0.839} \\
loao\_immune\_inflam & crossfit & normalized & 874 & 0.838 \ci{0.806}{0.868} & 0.600 & 0.158 & 0.835 \ci{0.827}{0.840} \\
loao\_infectious & split & constant & 727 & 0.912 \ci{0.885}{0.938} & 0.621 & 0.125 & 0.907 \ci{0.906}{0.909} \\
loao\_infectious & split & normalized & 727 & 0.856 \ci{0.826}{0.885} & 0.471 & 0.125 & 0.859 \ci{0.856}{0.861} \\
loao\_infectious & crossfit & constant & 727 & 0.883 \ci{0.852}{0.913} & 0.571 & 0.126 & 0.881 \ci{0.880}{0.882} \\
loao\_infectious & crossfit & normalized & 727 & 0.882 \ci{0.853}{0.912} & 0.539 & 0.126 & 0.882 \ci{0.879}{0.886} \\
loao\_neuro\_psych & split & constant & 253 & 0.909 \ci{0.868}{0.947} & 0.604 & 0.130 & 0.901 \ci{0.901}{0.901} \\
loao\_neuro\_psych & split & normalized & 253 & 0.877 \ci{0.830}{0.923} & 0.540 & 0.130 & 0.870 \ci{0.870}{0.877} \\
loao\_neuro\_psych & crossfit & constant & 253 & 0.913 \ci{0.869}{0.951} & 0.582 & 0.124 & 0.893 \ci{0.889}{0.897} \\
loao\_neuro\_psych & crossfit & normalized & 253 & 0.858 \ci{0.805}{0.906} & 0.535 & 0.124 & 0.858 \ci{0.858}{0.858} \\
loao\_oncology & split & constant & 450 & 0.951 \ci{0.931}{0.971} & 0.609 & 0.103 & 0.942 \ci{0.940}{0.947} \\
loao\_oncology & split & normalized & 450 & 0.942 \ci{0.919}{0.962} & 0.548 & 0.103 & 0.940 \ci{0.940}{0.942} \\
loao\_oncology & crossfit & constant & 450 & 0.949 \ci{0.927}{0.971} & 0.565 & 0.099 & 0.942 \ci{0.938}{0.947} \\
loao\_oncology & crossfit & normalized & 450 & 0.911 \ci{0.884}{0.938} & 0.463 & 0.099 & 0.911 \ci{0.909}{0.911} \\
loao\_other & split & constant & 1456 & 0.824 \ci{0.801}{0.846} & 0.618 & 0.169 & 0.813 \ci{0.808}{0.816} \\
loao\_other & split & normalized & 1456 & 0.764 \ci{0.738}{0.787} & 0.467 & 0.169 & 0.762 \ci{0.755}{0.764} \\
loao\_other & crossfit & constant & 1456 & 0.804 \ci{0.778}{0.825} & 0.537 & 0.164 & 0.793 \ci{0.784}{0.796} \\
loao\_other & crossfit & normalized & 1456 & 0.771 \ci{0.747}{0.794} & 0.470 & 0.164 & 0.772 \ci{0.769}{0.775} \\
\bottomrule\end{tabular}}\end{table}

%% file: tables/op.tex
\begin{table}[h]\centering\small
\caption{Answer-or-abstain operating points for the registry model, cross-fitted recipe, v2.1 binary contrasts: answer the contrasts whose predicted error scale is below a threshold. \emph{Calibration} rows choose the threshold on out-of-fold calibration residuals (largest answered fraction whose calibration MAE does not exceed the target) and report the answered fractions and the realised MAE when that threshold is applied unchanged to the evaluated set. \emph{Evaluated (oracle)} rows choose the threshold on the evaluated set itself, as in the submitted version, and meet the target by construction. The split recipe is in Table~\ref{tab:opsplit}.}
\label{tab:op}
\resizebox{\linewidth}{!}{\begin{tabular}{llllcccc}\toprule regime & target MAE & threshold chosen on & answered (calibration) & answered (contrasts) & answered (trials) & realised MAE, answered & MAE if all answered \\ \midrule
in\_distribution & 0.10 & calibration & 0.678 & 0.756 & 0.813 & 0.106 & 0.124 \\
in\_distribution & 0.10 & evaluated (oracle) & -- & 0.658 & 0.708 & 0.100 & 0.124 \\
in\_distribution & 0.08 & calibration & 0.397 & 0.483 & 0.538 & 0.095 & 0.124 \\
in\_distribution & 0.08 & evaluated (oracle) & -- & 0.289 & 0.338 & 0.080 & 0.124 \\
temporal\_2023plus & 0.10 & calibration & 0.759 & 0.837 & 0.850 & 0.113 & 0.130 \\
temporal\_2023plus & 0.10 & evaluated (oracle) & -- & 0.574 & 0.606 & 0.100 & 0.130 \\
temporal\_2023plus & 0.08 & calibration & 0.459 & 0.532 & 0.562 & 0.100 & 0.130 \\
temporal\_2023plus & 0.08 & evaluated (oracle) & -- & 0.256 & 0.276 & 0.078 & 0.130 \\
\bottomrule\end{tabular}}\end{table}

%% file: tables/opsplit.tex
\begin{table}[h]\centering\small
\caption{Answer-or-abstain operating points, split recipe, v2.1. The split recipe fits its error-scale model to in-sample residuals of the proper training half, which are compressed towards zero; the fitted scale therefore sits at its floor value for most contrasts (0.00 to 0.75 of the evaluated set across regimes), ties cannot be broken, and where the tied floor group alone exceeds the target no calibration threshold answers any contrast (answered fraction 0). The recipe is kept for comparability with the submitted version; the cross-fitted recipe (Table~\ref{tab:op}) is the one the main text uses.}
\label{tab:opsplit}
\resizebox{\linewidth}{!}{\begin{tabular}{llllcccc}\toprule regime & target MAE & threshold chosen on & answered (calibration) & answered (contrasts) & answered (trials) & realised MAE, answered & MAE if all answered \\ \midrule
in\_distribution & 0.10 & calibration & 0.000 & 0.000 & 0.000 & -- & 0.138 \\
in\_distribution & 0.10 & evaluated (oracle) & -- & 0.009 & 0.011 & 0.047 & 0.138 \\
in\_distribution & 0.08 & calibration & 0.000 & 0.000 & 0.000 & -- & 0.138 \\
in\_distribution & 0.08 & evaluated (oracle) & -- & 0.009 & 0.011 & 0.047 & 0.138 \\
temporal\_2023plus & 0.10 & calibration & 0.336 & 0.468 & 0.493 & 0.111 & 0.136 \\
temporal\_2023plus & 0.10 & evaluated (oracle) & -- & 0.158 & 0.171 & 0.098 & 0.136 \\
temporal\_2023plus & 0.08 & calibration & 0.115 & 0.161 & 0.174 & 0.102 & 0.136 \\
temporal\_2023plus & 0.08 & evaluated (oracle) & -- & 0.129 & 0.142 & 0.077 & 0.136 \\
\bottomrule\end{tabular}}\end{table}

%% file: tables/lmarea.tex
\begin{table}[h]\centering\small
\caption{Language model (Claude Sonnet~5) by therapeutic area, v2.1 binary test split (878 contrasts / 655 trials). Protocol-only prompts (headline condition) and posted-record prompts. skill(\full) and skill(\blind) against the training mean; $\IS$ with 95\% trial-level bootstrap intervals (2{,}000 resamples). v2 values in the released CSV.}
\label{tab:lmarea}
\begin{tabular}{lrrccc}\toprule area & contrasts & trials & skill(\full) & skill(\blind) & $\IS$ \\ \midrule
\multicolumn{6}{l}{\emph{protocol-only prompts}} \\
cardiometabolic & 45 & 34 & 0.332 & 0.289 & 0.060 \ci{-0.309}{0.552} \\
immune/inflam & 138 & 88 & 0.401 & 0.077 & 0.352 \ci{0.092}{0.577} \\
infectious & 205 & 148 & 0.674 & 0.558 & 0.262 \ci{0.076}{0.430} \\
neuro/psych & 48 & 37 & 0.382 & 0.201 & 0.226 \ci{0.019}{0.440} \\
oncology & 89 & 81 & 0.563 & 0.470 & 0.174 \ci{-0.434}{0.484} \\
other & 353 & 267 & 0.593 & 0.537 & 0.119 \ci{-0.093}{0.297} \\
\midrule
\multicolumn{6}{l}{\emph{posted-record prompts}} \\
cardiometabolic & 45 & 34 & 0.473 & 0.198 & 0.343 \ci{-0.253}{0.627} \\
immune/inflam & 138 & 88 & 0.489 & 0.141 & 0.406 \ci{0.194}{0.608} \\
infectious & 205 & 148 & 0.690 & 0.552 & 0.307 \ci{0.088}{0.486} \\
neuro/psych & 48 & 37 & 0.300 & 0.266 & 0.046 \ci{-0.403}{0.407} \\
oncology & 89 & 81 & 0.637 & 0.492 & 0.287 \ci{-0.171}{0.543} \\
other & 353 & 267 & 0.665 & 0.585 & 0.193 \ci{-0.013}{0.365} \\
\bottomrule\end{tabular}\end{table}

%% file: tables/lmsel.tex
\begin{table}[h]\centering\small
\caption{Selective prediction, language model (Claude Sonnet~5) condition A on the v2.1 binary test split (878 contrasts), ranked by the model's stated interval width: MAE at answered fractions 0.10 / 0.25 / 0.50 / 0.75 / 1.00, protocol-only (headline) and posted-record prompts. v2 values in the released CSV.}
\label{tab:lmsel}
\begin{tabular}{lccccc}\toprule prompt & 0.10 & 0.25 & 0.50 & 0.75 & 1.00 \\ \midrule
protocol-only & 0.018 & 0.039 & 0.066 & 0.080 & 0.098 \\
posted-record & 0.013 & 0.038 & 0.056 & 0.077 & 0.091 \\
\bottomrule\end{tabular}\end{table}